\documentclass[twocolumn,aps,superscriptaddress,prL,floatfix,showpacs,longbibliography,nofootinbib]{revtex4-2}

\usepackage{graphicx}
\usepackage{dcolumn}
\usepackage{bm}

\usepackage{ifthen,amsmath,amssymb}
\usepackage{hyperref}
\usepackage{float}
\usepackage{aas_macros}
\usepackage[export]{adjustbox}
\usepackage{bookmark}
\usepackage{newtxtext,newtxmath}
\usepackage[dvipsnames,table]{xcolor}
\usepackage{soul}

\newcommand{\bvel}{\boldsymbol v}
\newcommand{\bn}{\boldsymbol n}

\newcommand{\BH}[1]{\textcolor{black}{{{#1}}}}

\begin{document}

\preprint{APS/123-QED}

\title{Evidence for large baryonic feedback at low and intermediate redshifts from \\ kinematic Sunyaev-Zel'dovich observations with ACT and DESI photometric galaxies}

\author{B. Hadzhiyska}
\email{boryanah@alumni.princeton.edu}
\affiliation{Miller Institute for Basic Research in Science, University of California, Berkeley, CA, 94720, USA}\affiliation{Physics Division, Lawrence Berkeley National Laboratory, Berkeley, CA 94720, USA}
\affiliation{Berkeley Center for Cosmological Physics, Department of Physics, University of California, Berkeley, CA 94720, USA}

\author{S. Ferraro}
\affiliation{Physics Division, Lawrence Berkeley National Laboratory, Berkeley, CA 94720, USA}
\affiliation{Berkeley Center for Cosmological Physics, Department of Physics, University of California, Berkeley, CA 94720, USA}


\author{B. Ried Guachalla}
\affiliation{Department of Physics, Stanford University, Stanford, CA, USA 94305-4085}
\affiliation{Kavli Institute for Particle Astrophysics and Cosmology, 382 Via Pueblo Mall Stanford, CA 94305-4060, USA}
\affiliation{SLAC National Accelerator Laboratory 2575 Sand Hill Road Menlo Park, California 94025, USA}

\author{E. Schaan}
\affiliation{Kavli Institute for Particle Astrophysics and Cosmology,
382 Via Pueblo Mall Stanford, CA 94305-4060, USA}
\affiliation{SLAC National Accelerator Laboratory 2575 Sand Hill Road Menlo Park, California 94025, USA}

\author{J. Aguilar}
\affiliation{Lawrence Berkeley National Laboratory, 1 Cyclotron Road, Berkeley, CA 94720, USA}
\author{S. Ahlen}
\affiliation{Physics Dept., Boston University, 590 Commonwealth Avenue, Boston, MA 02215, USA}

\author{N. Battaglia}
\affiliation{Department of Astronomy, Cornell University, Ithaca, NY 14853, USA}

\author{J. R. Bond}
\affiliation{Canadian Institute for Theoretical Astrophysics, Toronto, Ontario M5S 3H8, Canada}

\author{D. Brooks}
\affiliation{Department of Physics \& Astronomy, University College London, Gower Street, London, WC1E 6BT, UK}
\author{E. Calabrese}
\affiliation{School of Physics and Astronomy, Cardiff University, The Parade, Cardiff, Wales CF24 3AA, UK}

\author{S. K. Choi}
\affiliation{Department of Physics and Astronomy, University of California, Riverside, CA 92521, USA}

\author{T. Claybaugh}
\affiliation{Lawrence Berkeley National Laboratory, 1 Cyclotron Road, Berkeley, CA 94720, USA}
\author{W. R. Coulton}
\affiliation{Kavli Institute for Cosmology Cambridge, Madingley Road, Cambridge CB3 0HA, UK}\affiliation{DAMTP, Centre for Mathematical Sciences, University of Cambridge, Wilberforce Road, Cambridge CB3 OWA, UK}
\author{K. Dawson}
\affiliation{Department of Physics and Astronomy, The University of Utah, 115 South 1400 East, Salt Lake City, UT 84112, USA}

\author{M. Devlin}
\affiliation{Department of Physics and Astronomy, University of Pennsylvania, 209 South 33rd Street, Philadelphia, PA, USA 19104}

\author{B. Dey}
\affiliation{Department of Physics \& Astronomy and Pittsburgh Particle Physics, Astrophysics, and Cosmology Center (PITT PACC), University of Pittsburgh, 3941 O'Hara Street, Pittsburgh, PA 15260, USA}
\author{P. Doel}
\affiliation{Department of Physics \& Astronomy, University College London, Gower Street, London, WC1E 6BT, UK}
\author{A. J. Duivenvoorden}
\affiliation{Center for Computational Astrophysics, Flatiron Institute, New York, NY 10010, USA, Joseph Henry Laboratories of Physics, Jadwin Hall, Princeton University, Princeton, NJ, USA 08544}
\author{J. Dunkley}
\affiliation{Joseph Henry Laboratories of Physics, Jadwin Hall, Princeton University, Princeton, NJ, USA 08544}\affiliation{Department of Astrophysical Sciences, Princeton University, Peyton Hall, Princeton, NJ 08544, USA}

\author{G. S. Farren}
\affiliation{DAMTP, Centre for Mathematical Sciences, University of Cambridge, Wilberforce Road, Cambridge CB3 OWA, UK}\affiliation{Kavli Institute for Cosmology Cambridge, Madingley Road, Cambridge CB3 0HA, UK}

\author{A. Font-Ribera}
\affiliation{Department of Physics \& Astronomy, University College London, Gower Street, London, WC1E 6BT, UK}\affiliation{Institut de F\'{i}sica d’Altes Energies (IFAE), The Barcelona Institute of Science and Technology, Campus UAB, 08193 Bellaterra Barcelona, Spain}
\author{J. E. Forero-Romero}
\affiliation{Departamento de F\'isica, Universidad de los Andes, Cra. 1 No. 18A-10, Edificio Ip, CP 111711, Bogot\'a, Colombia}\affiliation{Observatorio Astron\'omico, Universidad de los Andes, Cra. 1 No. 18A-10, Edificio H, CP 111711 Bogot\'a, Colombia}

\author{P. A. Gallardo}
\affiliation{Kavli Institute for Cosmological Physics, University of Chicago, Chicago, IL, 60637, USA}

\author{E. Gaztañaga}
\affiliation{Institut d'Estudis Espacials de Catalunya (IEEC), 08034 Barcelona, Spain}\affiliation{Institute of Cosmology and Gravitation, University of Portsmouth, Dennis Sciama Building, Portsmouth, PO1 3FX, UK}\affiliation{Institute of Space Sciences, ICE-CSIC, Campus UAB, Carrer de Can Magrans s/n, 08913 Bellaterra, Barcelona, Spain}
\author{S. Gontcho Gontcho}
\affiliation{Lawrence Berkeley National Laboratory, 1 Cyclotron Road, Berkeley, CA 94720, USA}
\author{M. Gralla}
\affiliation{Steward Observatory/Department of Astronomy, University of Arizona, Tucson, AZ 85721 USA}
\author{L. Le Guillou}
\affiliation{Sorbonne Universit\'{e}, CNRS/IN2P3, Laboratoire de Physique Nucl\'{e}aire et de Hautes Energies (LPNHE), FR-75005 Paris, France}
\author{G. Gutierrez}
\affiliation{Fermi National Accelerator Laboratory, PO Box 500, Batavia, IL 60510, USA}
\author{J. Guy}
\affiliation{Lawrence Berkeley National Laboratory, 1 Cyclotron Road, Berkeley, CA 94720, USA}
\author{J.~C. Hill}
\affiliation{Department of Physics, Columbia University, New York, NY 10027, USA}
\author{R. Hlo\v{z}ek}
\affiliation{Department of Astronomy and Astrophysics, University of Toronto, 50 St. George Street, Toronto, ON M5S 3H4, Canada}\affiliation{Dunlap Institute for Astronomy and Astrophysics, University of Toronto, 50 St. George Street, Toronto, ON M5S 3H4, Canada}
\author{K. Honscheid}
\affiliation{Center for Cosmology and AstroParticle Physics, The Ohio State University, 191 West Woodruff Avenue, Columbus, OH 43210, USA}\affiliation{Department of Physics, The Ohio State University, 191 West Woodruff Avenue, Columbus, OH 43210, USA}\affiliation{The Ohio State University, Columbus, 43210 OH, USA}
\author{S. Juneau}
\affiliation{NSF NOIRLab, 950 N. Cherry Ave., Tucson, AZ 85719, USA}
\author{R. Kehoe}
\affiliation{Department of Physics, Southern Methodist University, 3215 Daniel Avenue, Dallas, TX 75275, USA}
\author{T. Kisner}
\affiliation{Lawrence Berkeley National Laboratory, 1 Cyclotron Road, Berkeley, CA 94720, USA}
\author{A. Kremin}
\affiliation{Lawrence Berkeley National Laboratory, 1 Cyclotron Road, Berkeley, CA 94720, USA}
\author{M. Landriau}
\affiliation{Lawrence Berkeley National Laboratory, 1 Cyclotron Road, Berkeley, CA 94720, USA}

\author{R. H. Liu}
\affiliation{Berkeley Center for Cosmological Physics, Department of Physics, University of California, Berkeley, CA 94720, USA}

\author{T. Louis}
\affiliation{Universit\'e Paris-Saclay, CNRS/IN2P3, IJCLab, 91405 Orsay, France}
\author{N. MacCrann}
\affiliation{DAMTP, Centre for Mathematical Sciences, University of Cambridge, Wilberforce Road, Cambridge CB3 OWA, UK}
\author{A. de Macorra}
\affiliation{Instituto de F\'{\i}sica, Universidad Nacional Aut\'{o}noma de M\'{e}xico,  Cd. de M\'{e}xico  C.P. 04510,  M\'{e}xico}
\author{M. Madhavacheril}
\affiliation{Department of Physics and Astronomy, University of Pennsylvania, 209 South 33rd Street, Philadelphia, PA, USA 19104}
\author{M. Manera}
\affiliation{Departament de F\'{i}sica, Serra H\'{u}nter, Universitat Aut\`{o}noma de Barcelona, 08193 Bellaterra (Barcelona), Spain}\affiliation{Institut de F\'{i}sica d’Altes Energies (IFAE), The Barcelona Institute of Science and Technology, Campus UAB, 08193 Bellaterra Barcelona, Spain}
\author{A. Meisner}
\affiliation{NSF NOIRLab, 950 N. Cherry Ave., Tucson, AZ 85719, USA}
\author{R. Miquel}
\affiliation{Instituci\'{o} Catalana de Recerca i Estudis Avan\c{c}ats, Passeig de Llu\'{\i}s Companys, 23, 08010 Barcelona, Spain}\affiliation{Institut de F\'{i}sica d’Altes Energies (IFAE), The Barcelona Institute of Science and Technology, Campus UAB, 08193 Bellaterra Barcelona, Spain}
\author{K. Moodley}
\affiliation{Astrophysics Research Centre, School of Mathematics, Statistics and Computer Science, University of KwaZulu-Natal, Durban 4001, South Africa}
\author{J. Moustakas}
\affiliation{Department of Physics and Astronomy, Siena College, 515 Loudon Road, Loudonville, NY 12211, USA}
\author{T. Mroczkowski}
\affiliation{European Southern Observatory, Karl-Schwarzschild-Str.\ 2, Garching 85748, Germany}

\author{S. Naess}
\affiliation{Institute for theoretical astrophysics, University of Oslo,
Norway}

\author{J.  Newman}
\affiliation{Department of Physics \& Astronomy and Pittsburgh Particle Physics, Astrophysics, and Cosmology Center (PITT PACC), University of Pittsburgh, 3941 O'Hara Street, Pittsburgh, PA 15260, USA}

\author{M. D. Niemack}
\affiliation{Cornell University Physics Department, Cornell University, Ithaca, NY 14853, USA; Cornell University Astronomy Department, Cornell University, Ithaca, NY 14853, USA}

\author{G. Niz}
\affiliation{Departamento de F\'{i}sica, Universidad de Guanajuato - DCI, C.P. 37150, Leon, Guanajuato, M\'{e}xico}\affiliation{Instituto Avanzado de Cosmolog\'{\i}a A.~C., San Marcos 11 - Atenas 202. Magdalena Contreras, 10720. Ciudad de M\'{e}xico, M\'{e}xico}
\author{L. Page}
\affiliation{Joseph Henry Laboratories of Physics, Jadwin Hall, Princeton University, Princeton, NJ, USA 08544}
\author{N. Palanque-Delabrouille}
\affiliation{IRFU, CEA, Universit\'{e} Paris-Saclay, F-91191 Gif-sur-Yvette, France}\affiliation{Lawrence Berkeley National Laboratory, 1 Cyclotron Road, Berkeley, CA 94720, USA}
\author{B. Partridge}
\affiliation{Department of Physics and Astronomy, Haverford College, Haverford PA 19041 USA}
\author{W. J. Percival}
\affiliation{Department of Physics and Astronomy, University of Waterloo, 200 University Ave W, Waterloo, ON N2L 3G1, Canada}\affiliation{Perimeter Institute for Theoretical Physics, 31 Caroline St. North, Waterloo, ON N2L 2Y5, Canada}\affiliation{Waterloo Centre for Astrophysics, University of Waterloo, 200 University Ave W, Waterloo, ON N2L 3G1, Canada}
\author{F. Prada}
\affiliation{Instituto de Astrof\'{i}sica de Andaluc\'{i}a (CSIC), Glorieta de la Astronom\'{i}a, s/n, E-18008 Granada, Spain}

\author{F. J. Qu}
\affiliation{DAMTP, Centre for Mathematical Sciences, University of Cambridge, Wilberforce Road, Cambridge CB3 OWA, UK}
\affiliation{Kavli Institute for Cosmology Cambridge, Madingley Road, Cambridge CB3 0HA, UK}\affiliation{Kavli Institute for Particle Astrophysics and Cosmology, 382 Via Pueblo Mall Stanford, CA 94305-4060, USA}

\author{G. Rossi}
\affiliation{Department of Physics and Astronomy, Sejong University, Seoul, 143-747, Korea}
\author{E. Sanchez}
\affiliation{CIEMAT, Avenida Complutense 40, E-28040 Madrid, Spain}
\author{D. Schlegel}
\affiliation{Lawrence Berkeley National Laboratory, 1 Cyclotron Road, Berkeley, CA 94720, USA}
\author{M. Schubnell}
\affiliation{Department of Physics, University of Michigan, Ann Arbor, MI 48109, USA}\affiliation{University of Michigan, Ann Arbor, MI 48109, USA}

\author{B. Sherwin}
\affiliation{DAMTP, Centre for Mathematical Sciences, University of Cambridge, Wilberforce Road, Cambridge CB3 OWA, UK}

\author{N. Sehgal}
\affiliation{Physics and Astronomy Department, Stony Brook University, Stony Brook, NY 11794}

\author{H. Seo}
\affiliation{Department of Physics \& Astronomy, Ohio University, Athens, OH 45701, USA}

\author{C. Sif\'on}
\affiliation{Instituto de F\'isica, Pontificia Universidad Cat\'olica de Valpara\'iso, Casilla 4059, Valpara\'iso, Chile}

\author{D. Spergel}
\affiliation{Center for Computational Astrophysics, Flatiron Institute, 162 5th Avenue, New York, NY 10010, USA}
\author{D. Sprayberry}
\affiliation{NSF NOIRLab, 950 N. Cherry Ave., Tucson, AZ 85719, USA}

\author{S. Staggs}
\affiliation{Joseph Henry Laboratories of Physics, Jadwin Hall, Princeton University, Princeton, NJ, USA 08544}

\author{G. Tarl\'{e}}
\affiliation{University of Michigan, Ann Arbor, MI 48109, USA}
\author{C. Vargas}
\affiliation{Instituto de Astrof\'isica and Centro de Astro-Ingenier\'ia, Facultad de F\'isica, Pontificia Universidad Cat\'olica de Chile, Av. Vicu\~na Mackenna 4860, 7820436 Macul, Santiago, Chile}

\author{E. M. Vavagiakis}
\affiliation{Department of Physics, Duke University, Durham, NC 27708, USA; Department of Physics, Cornell University, Ithaca, NY 14853, USA}

\author{B. A. Weaver}
\affiliation{NSF NOIRLab, 950 N. Cherry Ave., Tucson, AZ 85719, USA}
\author{E. J. Wollack}
\affiliation{NASA/Goddard Space Flight Center, Greenbelt, MD 20771, USA}

\author{R. Zhou}
\affiliation{Physics Division, Lawrence Berkeley National Laboratory, Berkeley, CA 94720, USA}

\author{H. Zou}
\affiliation{National Astronomical Observatories, Chinese Academy of Sciences, A20 Datun Rd., Chaoyang District, Beijing, 100012, P.R. China}


\date{\today}

\begin{abstract}
Recent advances in cosmological observations have provided an unprecedented opportunity to investigate the distribution of baryons relative to the underlying matter. In this work, we show that the gas is more extended than the dark matter, and the amount of baryonic feedback at $z \lesssim 1$ disfavors low-feedback models such as that of state-of-the-art hydrodynamical simulation IllustrisTNG compared with high-feedback models such as that of the original Illustris simulation. This has important implications for bridging the gap between theory and observations and understanding galaxy formation and evolution. Furthermore, a better grasp of the baryon-dark matter link is critical to future cosmological analyses, which are currently impeded by our limited knowledge of baryonic feedback. Here, we measure the kinematic Sunyaev-Zel’dovich (kSZ) effect from the Atacama Cosmology Telescope (ACT), stacked on the luminous red galaxy (LRG) sample of the Dark Energy Spectroscopic Instrument (DESI) imaging survey. This is the first analysis to use photometric redshifts for reconstructing galaxy velocities. Due to the large number of galaxies comprising the DESI imaging survey, this is the highest signal-to-noise stacked kSZ measurement to date: we detect the signal at 13$\sigma$, \BH{finding strong evidence that the gas is more spread out than the dark matter, as well as a preference for larger feedback compared to some commonly used state-of-the-art hydrodynamical simulations.} Our work opens up the possibility of recalibrating large hydrodynamical simulations using the kSZ effect. In addition, our findings point towards a way of alleviating inconsistencies between weak lensing surveys and cosmic microwave background (CMB) experiments, such as the `low $S_8$' tension, and shed light on long-standing enigmas in astrophysics, such as the `missing baryon' problem.
\end{abstract}

\maketitle


\section{Introduction}
\label{sec:intro}

Baryons, though comprising more than 15\% of the universe's total matter content, continue to elude precise mapping in relation to the underlying dark matter \cite{Fukugita:2004ee}. This poses a significant challenge for the next generation of large-scale structure experiments, especially those measuring weak lensing, including the Vera Rubin Observatory \cite{2012arXiv1211.0310L,2019ApJ...873..111I}, Euclid \cite{2013LRR....16....6A}, and the Nancy Grace Roman Space Telescope \cite{2015arXiv150303757S}.

Realizing the full potential of these surveys necessitates sub-percent-level knowledge of baryonic effects on cosmological observables. Furthermore, unraveling the astrophysical processes governing baryons within galaxies and galaxy clusters holds the key to deciphering the mysteries of galaxy formation and evolution. The bulk of baryons reside in the circumgalactic medium (CGM) and the intracluster medium (ICM), the baryon abundance and composition of which is shaped by phenomena such as active galactic nuclei (AGN) winds and supernova explosions \citep{Cen_2006}.

Among the most potent tools for probing the elusive baryon distribution is the measurement of the Sunyaev-Zel'dovich (SZ) effect around galaxies. Stemming from the interaction of free electrons in the CGM and ICM with cosmic microwave background (CMB) photons, the SZ effect provides a window into the thermodynamic properties of galaxy groups and clusters, critical for resolving a multitude of astrophysical and cosmological enigmas. In particular, the thermal Sunyaev-Zel'dovich (tSZ) effect arises from the inverse Compton scattering of CMB photons by the hot thermal gas within groups and clusters. Its magnitude, directly proportional to the electron pressure integrated along the line-of-sight, offers insights into the thermodynamic conditions of these cosmic structures. In contrast, the kinematic Sunyaev-Zel'dovich (kSZ) effect results from the encounters between CMB photons and free electrons in bulk motion relative to the CMB rest-frame. This effect depends on the integrated electron density along the line-of-sight, multiplied by the peculiar velocity making it a powerful tool for tracing the spatial distribution of baryons, even to the outskirts of galaxies and clusters (e.g., see \citep{1999PhR...310...97B,2019SSRv..215...17M} for a review on the SZ effect). In particular, the kSZ effect can be used to measure the gas density in the outskirts of galaxies, independent of any assumptions on temperature or metallicity.

\BH{In this work, we measure the baryon profiles around Dark Energy Spectroscopic Instrument (DESI) luminous red galaxies (LRGs) using the stacking of the kSZ effect. The kSZ effect has been measured using a variety of techniques over the years, including the first detections using the pairwise estimator \citep{2012PhRvL.109d1101H, 2016MNRAS.461.3172S}, through `velocity reconstruction' templates (and the closely related `velocity-inversion') \citep{ACTPol:2015teu, Schaan21, 2015PhRvL.115s1301H, 2016A&A...586A.140P, 2021A&A...645A.112T}, cross-correlations with redshift fluctuation maps \citep{2021MNRAS.503.1798C}, and projected fields techniques \citep{2016PhRvL.117e1301H,2016PhRvD..94l3526F}, among others.} Our findings shed light on the complex relationship between baryonic matter and dark matter. The novelty of this work lies in the following three aspects: 1) its pioneering use of photometric redshifts for probing the stacked kSZ effect, which departs from the traditional use of spectroscopic redshifts; 2) challenging the predictions of state-of-the-art hydrodynamical simulations -- we find that the feedback processes appear to be stronger in the real Universe than in simulations -- in other words, the baryons are expelled more violently from the cores of galaxy groups and clusters; 3) its implications for alleviating or resolving discrepancies in weak lensing analysis;
most notably, the `low $S_8$' tension found when comparing the inferred values of the clumpiness parameter, $S_8 \equiv \sigma_8 (\Omega_m/0.3)^{1/2}$, from weak lensing with CMB probes (e.g., see \citep{2022JHEAp..34...49A} and references therein), and the `Lensing is low' tension, which constitutes a discrepancy between galaxy clustering and galaxy-galaxy lensing \citep{2017MNRAS.467.3024L,2023MNRAS.518..477A}.

This paper is organized as follows. Section~\ref{sec:data} and Section~\ref{sec:methods} offer a description of our data and methodology, respectively, while Section~\ref{sec:res} describes our results. Section~\ref{sec:conc} discusses the broader implications of our findings for both astrophysics and cosmology, particularly their potential for addressing the $S_8$ and the `Lensing is low' discrepancies.

\section{Data}
\label{sec:data}

\subsection{DESI imaging survey} 
\label{sec:desi}

The Dark Energy Spectroscopic Instrument (DESI) is a robotic, fiber-fed, highly multiplexed spectroscopic surveyor that operates on the Mayall 4-meter telescope at Kitt Peak National Observatory \citep{2022AJ....164..207D}. DESI, which can obtain simultaneous spectra of almost 5000 objects over a $\sim$$3^\circ$ field \citep{2016arXiv161100037D,2023AJ....165....9S,2023arXiv230606310M}, is currently conducting a five-year dark energy survey of about a third of the sky \citep{2013arXiv1308.0847L}. This campaign will obtain spectra for approximately 40 million galaxies and quasars \citep{2016arXiv161100036D}.


In this paper, we use the photometric sample of DESI Luminous Red Galaxies (LRGs) and a higher-density extended LRG sample, referred to as `Main LRGs' and `Extended LRGs' \cite{2017PASP..129f4101Z,2019AJ....157..168D,2020RNAAS...4..181Z,2023AJ....165...58Z,Zhou:2023gji}. {The parent imaging is the DESI Legacy Imaging Survey, which is a combination of optical and mid-infrared imaging used for DESI target selection from three telescopes: the Dark Energy Camera Legacy Survey (DECaLS), the Mayall $z$-band Legacy Survey (MzLS), and the Beijing–Arizona Sky Survey (BASS).} We also make use of the reliable photometric redshifts presented in \citet{Zhou:2023gji} for Data Release 9 (DR9) and Data Release 10 (DR10), which have been calibrated via DESI spectroscopic redshifts and thus have only a small fraction of contaminants and a mean error of $\sigma_z/(1+z) \lesssim 0.02$. \BH{For DR10, we adopt the $i$-band $z$ estimates (\texttt{Z\_PHOT\_MEDIAN\_I}), as they are more accurate.} Similarly to \citet{Zhou:2023gji}, we define four tomographic bins: [0.4, 0.54], [0.54 0.713], [0.713, 0.86] and [0.86 1.024]
 (see Table~\ref{tab:chi2} for mean bin redshifts and number of galaxies in the overlap with ACT). 
We take into account the estimated error in the photometric redshifts for each galaxy, by imposing a maximum cut on the estimated error of $\sigma_{z,{\rm max}} = 0.05 (1+z)$. This removes $\sim$3\% of the galaxies, including outliers with anomalously large redshift errors that could otherwise bias the reconstructed velocities by diluting the density signal \citep[see][for tests on simulations]{2024PhRvD.109j3534H}. In App.~\ref{app:corr_dr9}, we investigate the effects of not imposing such a cut and find that, for the most part, they are negligible. 


\subsection{ACT temperature maps} 
\label{sec:act}

\BH{This paper employs the CMB maps
maps from the Data Release 6 (DR6) of the Atacama Cosmology Telescope (ACT) \cite{2020JCAP...12..046N}, a 6 m telescope that was located in the Atacama Desert of Chile and measured the CMB from 2007 to 2022. The multifrequency observational program of DR6 targeted the `wide' field; for this work, we use only the night-time portion of the data taken in the first five observing seasons (2017-2021). In particular, we make use of the harmonic-space Internal Linear Combination (hILC) 
maps \cite{ACT:2023wcq}, which are constructed by combining the high-resolution ($\sim$1.5 arcmin) DR6 maps multifrequency observations from 2017 to 2022 at three frequency bands: f090 (77–112 GHz), f150 (124–172 GHz), and f220 (182–277 GHz), alongside \textit{Planck} data ($\sim$ 5 arcmin resolution) on large scales \citep{2016ApJS..227...21T,2024ApJ...962..113M}. The resulting hILC map covers roughly a third of the sky (32\%), with a resolution (i.e., effective beam size Full-Width at Half Maximum) of 1.6 arcmin and mean white noise level in the temperature of approximately 15 $\mu $K-arcmin}.
The ACT maps are produced in the plate-carr\'{e}e (CAR) projection scheme. This analysis uses the first version of the ACT DR6 maps, dr6.01.
Similarly to \cite{Schaan21}, we apply a mask on the ACT map that removes all detected clusters \citep{2024ApJ...962..112Q} and point sources as well as $5\sigma$ outliers in any of the filtered temperature bins, since those can bias the stacked profiles.

\section{Methods}
\label{sec:methods}

\subsection{Reconstruction}
\label{sec:rec}

If we naively stacked the kSZ signal around a sample of galaxies, the effect would cancel out, as each object has an equal chance of moving towards or away from the observer. Therefore, to selectively extract the kSZ effect from CMB maps, we employ an estimate of the peculiar velocity of each galaxy in the line-of-sight direction, reconstructed from the three-dimensional galaxy overdensity field. In particular, we can obtain an estimate of the line-of-sight velocity field by solving the linearized continuity equation in redshift space \citep{Padmanabhan12}, similarly to the reconstruction method applied in Baryon Acoustic Oscillations (BAO) analysis:
\begin{equation}
    \nabla \cdot \bvel + \frac{f}{b} \nabla \cdot [(\bvel \cdot \hat \bn) \hat \bn] = -a H f \frac{\delta_g}{b} ,
\end{equation}
where $\delta_g$ is the observed galaxy overdensity field, $\bvel$ is the peculiar velocity field, $\hat \bn$ is the line-of-sight unit vector, $H(z)$ is the redshift-dependent Hubble parameter, $f$ is the logarithmic growth rate, defined as $f \equiv d \ln(D)/d \ln(a)$ with $D(a)$ the growth factor and $a$ the scale factor. Here, we assume that the galaxy overdensity $\delta_g$ is related to the matter overdensity, $\delta$, by a linear bias factor, $b$, such that $\delta_g = b \delta$. 

We construct the galaxy overdensity field separately in each tomographic bin.
Due to the difference in the depth of the three regions that make up our sample, designated `DES', `N' and `S', we generate the sample of randoms for each region before joining them back together to evaluate the galaxy overdensity field in the given tomographic bin.

In order to obtain an estimate of the peculiar velocity of each galaxy, we adopt the standard BAO reconstruction method of \citep{ESSS07}, implemented in the package \texttt{pyrecon} \footnote{\url{https://github.com/cosmodesi/pyrecon}} \citep{2015MNRAS.450.3822W}. This method yields an estimate of the first-order galaxy displacement field, which can be converted into an estimate for the velocity. A study performed on realistic DESI-like light cone mocks \cite{2024PhRvD.109j3534H} informs us that the cross-correlation coefficient,
$\smash{r \equiv \langle v_{||}^{\rm halo} v_{||}^{\rm rec} \rangle / {\sigma_{||}^{\rm halo} \sigma_{||}^{\rm rec}}}$,
between the reconstructed galaxy velocities along the line-of-sight (denoted by subscript $||$) and the host halo velocities, which captures imperfections in the velocity reconstruction, is about $r \approx 0.3$ (with an uncertainty of about 10\%), the value we adopt in this study, while that of the spectroscopic sample is about $r \approx 0.64$. In other words, the reconstruction performance deteriorates by about half when using photometric redshifts, a prediction also confirmed by a similar study in the `snapshot geometry' \cite{2024PhRvD.109j3533R}. 
Nonetheless, the measurement around the photometric LRG sample yields a high signal-to-noise ratio (SNR) due to the large number of galaxies comprising the imaging survey. We note that LRGs are found mostly in galaxy groups with a mean linear bias of $b \approx 2.2$ \citep{Yuan:2023ezi}.
In addition to the cuts in \cite{Schaan21}, for our fiducial analysis, 
\BH{we remove outliers in the reconstructed velocities: in particular, we ensure that the number of galaxies in each velocity bin is symmetric around the mean by random downsampling}, which avoids unwanted biases from massive clusters and guarantees that in the absence of a kSZ signal, our estimator yields zero mean signal. In App.~\ref{app:corr_dr9}, we show that these corrections have a mostly negligible effect: the uncorrected measurements tend to have marginally higher SNR (as a result of the larger number of galaxies in the samples) except for bin 3, which has a lower SNR due to its slightly asymmetric line-of-sight velocity distribution that improves once we remove the outliers. 

\subsection{Estimator}
\label{sec:est}

Once we have estimates of the velocities, we can measure the CMB temperatures $\mathcal{T}_i(\theta_d)$ around each galaxy $i$ using a compensated aperture photometry (CAP) filter \citep{Schaan21}.
There are several benefits to using the CAP filter. Unlike a matched filter, the CAP filter is agnostic about the profile shape and because it is measured at different radii, it allows us to reconstruct the spherically averaged profiles. 
It is very effective at removing the large-scale primary CMB, as well as the uncorrelated part of the signal, leaving, in principle,  only the sum of the one- and two-halo terms \citep[see][for tests on simulations]{2023MNRAS.526..369H}. Additionally, the CAP-filtered profiles behave similarly to a cumulative density profile for large radii. The CAP filter is defined as:
\begin{equation}
\label{eq:ap}
\mathcal{T}(\theta_{d})=
\int {\rm d}^2\theta \, \delta T(\theta) \, W_{\theta_{d}}(\theta) \,.
\end{equation}
where $\delta T(\theta)$ are the CMB temperature fluctuations and the filter $W_{\theta_{d}}$ is chosen as:
\begin{equation}
W_{\theta_{d}}(\theta) =
\left\{
\begin{aligned}
1& &  &\text{for} \, \theta < \theta_d \,, \\
-1& &  &\text{for} \, \theta_d \leq \theta \leq \sqrt{2}\theta_d \,, \\
0& & &\text{otherwise}. \\
\end{aligned}
\right.
\end{equation}
Thus, we add the integrated temperature fluctuation in a disk with radius $\theta_d$ and subtract a concentric ring of the same area as the disk, at each radial bin. Similarly to \cite{Schaan21}, we choose nine radial bins, $\theta_d$, spanning between 1 and 6 arcmin.
For the four redshift bins (see Table~\ref{tab:chi2}),
this range corresponds roughly to a physical range of (0.36, 2.2), (0.42, 2.5), (0.46, 2.8) and (0.48, 2.9) Mpc, respectively.

We adopt the velocity-weighted, uniform-mean estimator from \cite{Schaan21}:
\begin{equation}
    \hat{T}_{\rm kSZ}(\theta_d) = -
    \frac{1}{r}
    \frac{v_{\rm rms}^{\rm rec}}{c}
    \frac{\sum_i \mathcal{T}_i(\theta_d) (v_{{\rm rec}, i}/c)}{\sum_i (v_{{\rm rec}, i}/c)^2},
    \label{eq:kSZ_est}
\end{equation}
where $v_{\rm rms}^{\rm rec}$ is the rms of the radial component of the reconstructed velocities, $v_{{\rm rec}, i}$, $c$ is the speed of light, and $r$ is the cross-correlation coefficient between the reconstructed and true velocity, evaluated from mock simulations.
We use the publicly available pipeline \texttt{ThumbStack}\footnote{\url{https://github.com/EmmanuelSchaan/ThumbStack}} to apply the estimator to the DESI and ACT data.

To interpret the kSZ profiles obtained through the above method, we need to estimate the covariance of the stacked measurements. We do so by applying a bootstrap resampling to the signal at each galaxy location. In particular, we draw 10,000 realizations of the galaxy catalogs (with repetition), and infer the covariance matrices from the scatter across the resampled stacked profiles (see Section~\ref{app:cov}).

\section{Results}
\label{sec:res}

\subsection{Stacked kSZ signal per redshift bin}

\begin{figure}[h]
    \centering
    \includegraphics[width=0.45\textwidth]{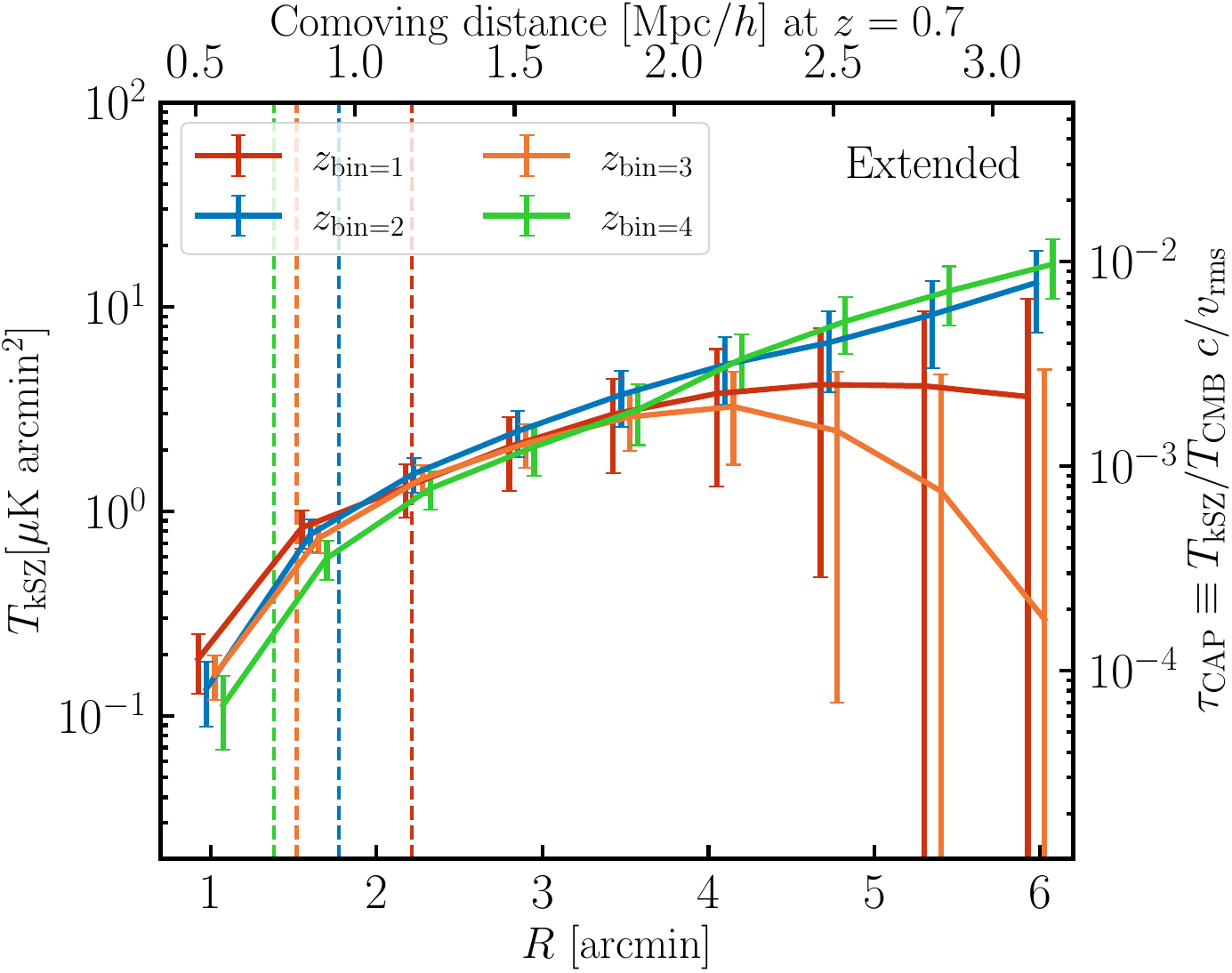}
    \includegraphics[width=0.45\textwidth]{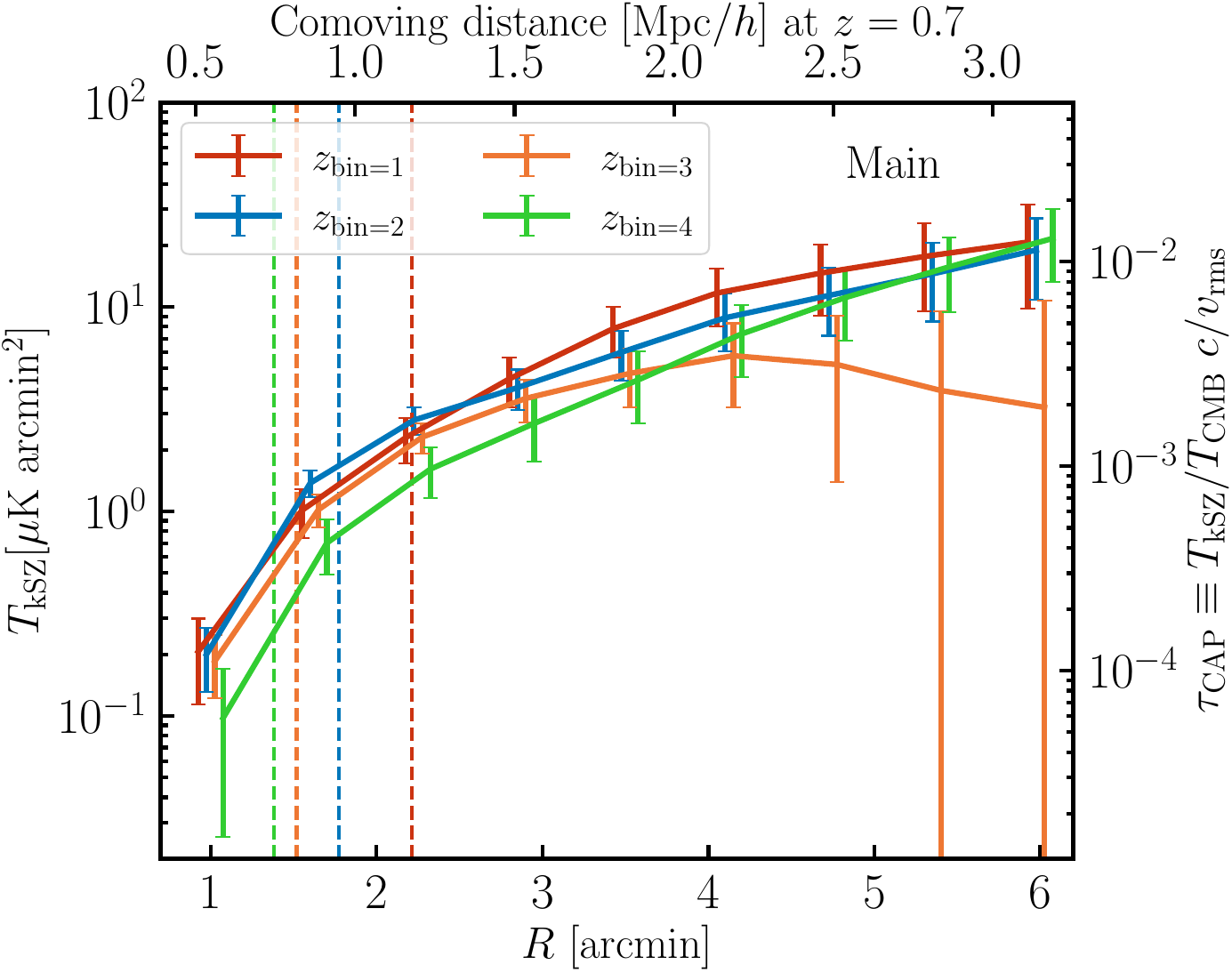}
    \caption{Stacked kSZ signal in each redshift bin of the photometric LRG sample (\textit{top:} Extended, \textit{bottom:} Main) as a function of radius of the CAP filter, which approximately corresponds to the distance from the galaxy group center associated with the LRGs. The kSZ signal is obtained by stacking the hILC ACT DR6 map (with an effective ${\rm FWHM} = 1.6'$) and is detected at $> 10\sigma$ relative to dark matter in each bin
    (see Table~\ref{tab:chi2}). The vertical dashed line indicates the mean virial radius of the host halos for each bin. We show the full covariance of the data in App.~\ref{app:cov} and note that the points at large aperture are significantly correlated. We see that the gas profile is well extended beyond the virial radius, suggesting that the feedback activity in LRG halos is strong enough to push much of the gas away from the galaxy group center. The top $x$ axis and the right $y$ axis show the comoving distance and $\tau_{\rm CAP}$ at the mean redshift.
    }
    \label{fig:kSZ}
\end{figure}

    \begin{table}
        \centering
        {\renewcommand{\arraystretch}{1.5}
        \begin{tabular}{l||c|c|c|c}
            \hline
            \hline
            LRG sample & \# of galaxies & $z_{\rm mean}$ & $\chi^2_{\rm null}$ & ${\rm SNR}_{\rm null}$ \\[2pt] 
            \hline
        Extended DR9 $z_{\rm bin=1}$ & 963,631 & 0.47 & 43.2 & 6.4 \\[-0.5pt]
        Extended DR9 $z_{\rm bin=2}$ & 1,658,313 & 0.63 & 69.4 & 7.1 \\[-0.5pt]
        Extended DR10 $z_{\rm bin=3}$ & 1,951,646 & 0.79 & 80.3 & 8.2 \\[-0.5pt]
        Extended DR10 $z_{\rm bin=4}$ & 1,690,171 & 0.92 & 34.0 & 4.6 \\[-0.5pt]
        \hline
        Extended DR9+10 all & 6,850,072 & 0.75 & 203.0 & 13.5 \\[-0.5pt]
        \hline
        Main DR9 $z_{\rm bin=1}$ & 422,350 & 0.47 & 25.2 & 4.4 \\[-0.5pt]
        Main DR9 $z_{\rm bin=2}$ & 795,393 & 0.63 & 78.8 & 7.8 \\[-0.5pt]
        Main DR10 $z_{\rm bin=3}$ & 753,945 & 0.79 & 65.9 & 7.3 \\[-0.5pt]
        Main DR10 $z_{\rm bin=4}$ & 629,367 & 0.93 & 20.8 & 3.2 \\[-0.5pt]
        \hline
        Main DR9+10 all & 2,882,904 & 0.74 & 166.7 & 12.1 \\[-0.5pt]
        \hline
        \hline
        \end{tabular}%
        }
        \caption{Statistics of the detection for each redshift bin and LRG sample in terms of the null $\chi^2_{\rm null}$, the SNR with respect to null, defined as $\smash{{\rm SNR}_{\rm null} \equiv (\chi^2_{\rm null} - \chi^2_{\rm bf})^{\frac{1}{2}}}$, with best-fit coming from the Illustris-1 simulation curves at $z = 0.5$ (see Fig.~\ref{fig:kSZ_sim}). We detect the signal at 13.5$\sigma$ and 12.1$\sigma$ for the Extended and Main sample, respectively, with nine degrees of freedom.
        }
        \label{tab:chi2}
    \end{table}

Fig.~\ref{fig:kSZ} shows the stacked kSZ signal (Eq.~\ref{eq:kSZ_est}) from ACT for the four photometric bins of the DR10 Extended and Main LRG samples obtained from the DESI Imaging Survey (see Section~\ref{sec:desi}). The signal is shown as a function of CAP filter radius and detected at a high significance. We note that the error bars at large aperture are highly correlated (see App.~\ref{app:cov} on the covariance estimation). To aid interpretation, we convert the $x$ and $y$ axis into proper distance and CAP-filtered optical depth ($\tau_{\rm CAP} = T_{\rm kSZ}/T_{\rm CMB} \  c/v_{\rm rms}$), respectively, calculated at the mean redshift, $z \approx 0.7$. We note that the optical depth measures the integrated gas density along the line-of-sight, 
\begin{equation}
    \tau (z) \equiv \int n_e (\chi \hat  \bn, z) \sigma_T \frac{{\rm d} \chi}{1+z} ,
\end{equation}
where $\sigma_T$ is the Thomson scattering cross section, $\chi$ is the comoving distance to redshift $z$, and $n_e$ is the free electron physical number density.

In Table~\ref{tab:chi2}, we show the significance of detection and corresponding SNR and chi-squared values for each of the four bins as well as the combination of all four bins. We define the SNR with respect to null\footnote{Corresponding to the inverse of the fractional error on the fit of a single amplitude parameter to the data. This often called the ``detection significance'' in number of Gaussian standard deviations $\sigma$.} as $\smash{{\rm SNR}_{\rm null} \equiv (\chi^2_{\rm null} - \chi^2_{\rm bf})^{\frac{1}{2}}}$, with best-fit coming from the Illustris-1 simulation\footnote{\url{https://www.tng-project.org/data/}} curve at $z = 0.5$ (see Fig.~\ref{fig:kSZ_sim} for more details) with a free parameter for the amplitude. We detect the signal at $\sim$13$\sigma$ as shown in Table~\ref{tab:chi2}.


Here we have defined the $\chi^2$ metric in the standard way using the covariance matrix $C$ (see App.~\ref{app:cov}):
\begin{equation}
    \chi^2_{\rm model} \equiv (D - M)^T C^{-1} (D - M) \ ,
\end{equation}
where $D$ is our stacked kSZ measurement (data vector), and $M$ is the model we are comparing against (if null, $M = 0$).

To take advantage of the larger number of objects in DR9 and their relatively small photometric $z$ errors at low $z$ (compared with high $z$), we quote the DR9 results for bin 1 and 2 and the DR10 results for bin 3 and 4. For the combined `DR9+10 all,' we use DR10 reconstructed velocities where available and otherwise DR9 ones in order to maximize the signal-to-noise. We provide detailed comparison between DR9 and DR10 in Tables~\ref{tab:chi2_corr} and~\ref{tab:chi2_corr_dr9} 
finding a high level of congruence between the two.
Compared with the previous measurement using CMASS LRGs and ACT DR4 \citep{Schaan21}, $\chi^2_{\rm null} \approx 86$ (same number of radial bins), we see that the total chi-squared for our Extended and Main samples is $\chi^2_{\rm null} \approx 200$ and 170, respectively,
or about twice higher.
In App.~\ref{app:cmass}, we find our curves to be in excellent agreement with \citep{Schaan21}.

Intuitively, one can think of these curves as showing roughly the cumulative gas density distribution. At large radii, we expect the profiles to become shallower as more and more of the gas is enclosed. The gas profiles are steep beyond the virial radius, suggesting that a significant fraction of the gas resides beyond it (see discussion of Fig.~\ref{fig:kSZ_sim}, where we test this conjecture). The decline in amplitude at large apertures of bin 3 is due to noise from the primary CMB and the larger fraction of photometric $z$ outliers in that bin (see App.~\ref{app:corr_dr9}). 

\subsection{Comparison with simulations}

\begin{figure}
    \centering
    \includegraphics[width=0.5\textwidth]{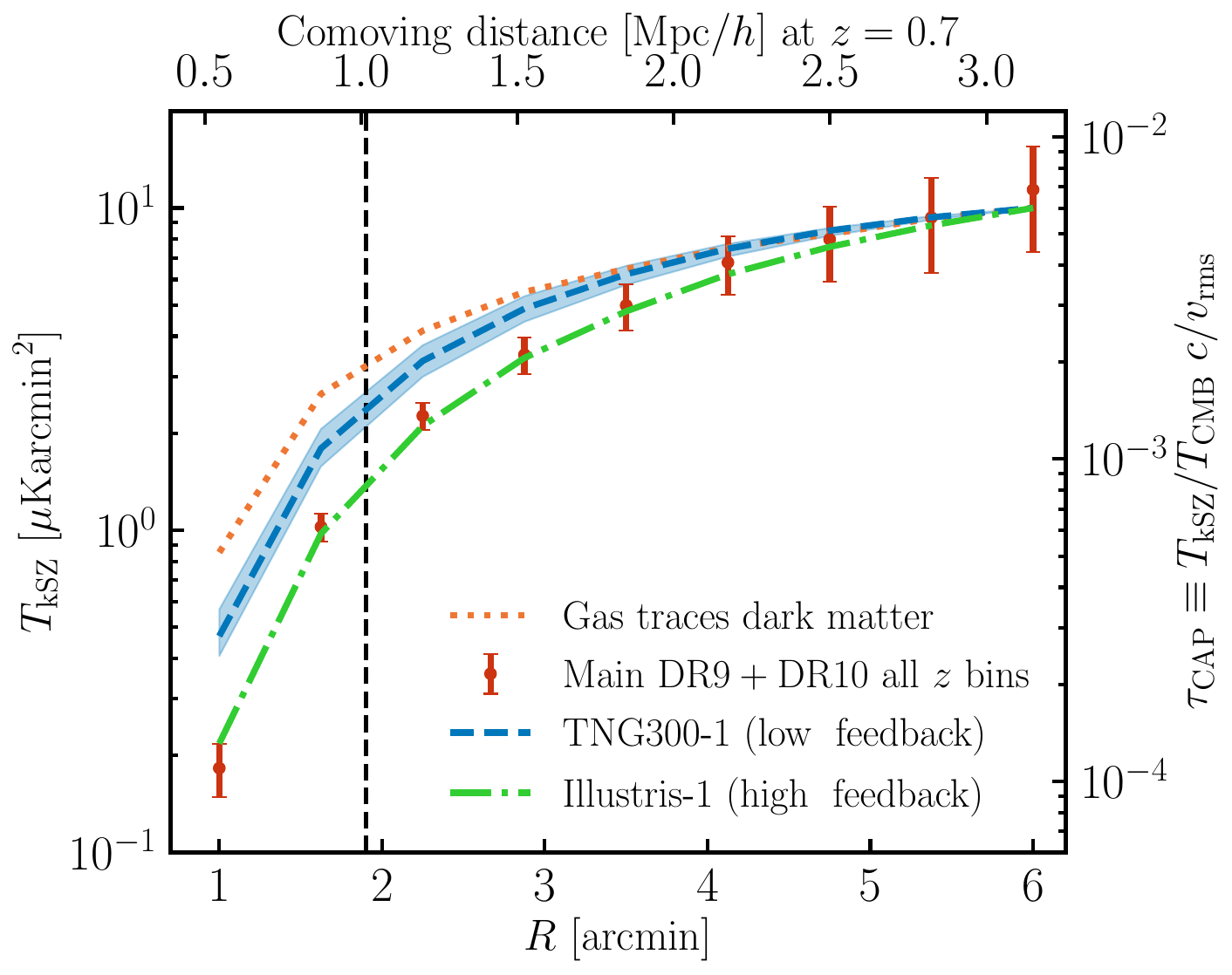}
    \caption{Comparison between the kSZ signal from the combined Main sample from all redshift bins (red error bars) and the modeled kSZ signal from the TNG and Illustris simulations. These curves approximate the cumulative gas profiles at large radii. The right $y$ axis converts the kSZ signal into a measure of the optical depth $\tau$, i.e., the integrated gas density along the line-of-sight, via $\smash{\tau = T_{\rm kSZ}/T_{\rm CMB} c/v_{\rm rms}}$. The top $x$ axis shows the gas profiles as a function of comoving distance from the center of DESI groups. We find at high a significance 
    that the gas is more extended than the dark matter (red dotted curve). 
    The gap between the data and the TNG curve (blue dashed curve) indicates that the data disfavor prescriptions of weak baryonic feedback in simulations. In contrast with TNG, the old Illustris model (green dash-dotted curve) appears much more consistent with the data, implying that models with large baryonic feedback are preferred. \BH{To visually aid the comparison, we have normalized all simulations to enclose the same gas mass at large aperture: we discuss a more quantitative comparison and related caveats in the main text.} The vertical line shows the mean virial radius for this sample. The thin blue band around the TNG curve represents the range of alternative scenarios considered (see the text). We show the full covariance of the data in App.~\ref{app:cov}.} 
    \label{fig:kSZ_sim}
\end{figure}
\BH{We first compare our data to dark matter simulations: the dark matter profiles are obtained by stacking on dark-matter-only maps computed using the TNG300-1-Dark simulation (dark-matter-only counterpart to TNG300-1), with the results shown in Fig. \ref{fig:kSZ_sim}. }

\BH{Quantifying the level of agreement or disagreement between data and simulations without the use of external measurements or priors is challenging because the amplitude of the results is very sensitive to the mass of the sample and to finite-box effects in simulations. The same selection applied to different simulations will result in different mean masses, complicating the comparison.}

\BH{Similarly to \citep{RiedGuachalla:2025byu}, we adopt two methods, each with advantages and drawbacks. In the first, we normalize all simulations to contain the same gas mass at large aperture (5 arcmin in this case), matching the value of the best-fitting (lowest $\chi^2$ with the data) simulation, in this case Illustris-1. This ensures that we are comparing halos of the same mass across all simulations, and that such gas mass is compatible with the data. We call this ``fixed amplitude''. Then we quantify the relative difference with respect to the best-fit simulation as $\smash{{\rm SNR}_{\rm DM}^{\rm fixed} \equiv (\chi^2_{\rm DM} - \chi^2_{\rm bf})^{\frac{1}{2}}}$, where the DM amplitude has been fixed as described. We find $\smash{{\rm SNR}_{\rm DM}^{\rm fixed}} = 39$ and 43, for the Main and Extended samples, respectively. We note that a major drawback of this procedure is that, while it ensures consistency of gas mass between the data and the simulations, it doesn't properly marginalize over the uncertainty in this quantity, resulting in larger values of $\smash{{\rm SNR}_{\rm DM}}$.}

\BH{Alternatively, we can let the amplitude of each simulation be completely free and fit it to the data: this fully marginalizes over the uncertainty in all of the data points. The drawback is that it is allowed to find best-fit solutions with very low gas or halo mass, potentially ruled out by external data (for example, weak lensing or fits to the galaxy correlation function). Nonetheless, this is the most conservative comparison we can perform without joint analysis of external data, and therefore we'll take it as the baseline for our conclusions. We can define a ``free amplitude'' $\smash{{\rm SNR}_{\rm DM}^{\rm free} \equiv (\chi^2_{\rm DM} - \chi^2_{\rm bf})^{\frac{1}{2}}}$, where the DM curve has the best-fit amplitude. For both samples, we find $\smash{{\rm SNR}_{\rm DM}^{\rm free}} \approx 6$. Given that even in the more conservative case of ``free amplitude'', the dark matter curve is a much poorer fit to the data, we conclude that we find strong evidence that the gas as measured by the kSZ effect, is significantly more extended than the dark matter.}

\BH{We note that the limitations of both approaches can be mitigated by the inclusion of external data such as weak lensing or galaxy clustering (among others), which independently constrain the mass of the sample \citep{McCarthy:2024tvp}. Such joint analyses will be able to sharpen the conclusions significantly, but they are beyond the scope of this paper and will be the subject of future work.}

\BH{We then} study the comparison between the state-of-the-art hydrodynamical simulations IllustrisTNG-300-1 (TNG300-1) \citep{2019ComAC...6....2N} and the measured gas profiles from the data. We also show a comparison with the older model of the Illustris-1 simulation \citep{2015A&C....13...12N}, which has known shortcomings with predicting observations such as galaxy morphologies and colors, various cluster properties, and gas fractions
\citep{2014MNRAS.445..175G,2019MNRAS.486.4686K,2019MNRAS.483.4140R,2019ComAC...6....2N}. 
\BH{We provide details on how the measurements on simulations are made in App.~\ref{app:sims}.} In App.~\ref{app:mass}, we split the galaxies into stellar mass bins estimated in \citep{2023AJ....165...58Z} and detect the mass evolution of the signal at high significance.

We perform a number of tests on both hydrodynamical simulations to ensure that the gap we see is significant: 
we vary the satellite fraction from 0\% to 30\%\footnote{Centrals are selected as the most massive subhalo in each host halo. To vary the satellite fraction, we randomly discard centrals or satellites.} (note that the constraints from \citep{Yuan:2023ezi} are $11 \pm 1\%$), we put all galaxies at the centers of their host halos, we vary the number density (between half and twice the fiducial value, i.e. from $3 \times 10^{-4}$ to $1 \times 10^{-3} \ [{\rm Mpc}/h]^{-3}$) and thus stellar mass threshold of the extracted LRGs, and we add noise to the halo velocities. 
In Fig.~\ref{fig:kSZ_sim}, we show the default scenario for both simulations, and in shaded color for TNG300-1, we display the minimum and maximum deviations from the default caused by all the aforementioned alternative scenarios. 
Despite considering scenarios that push the bounds of physically reasonable models, we see that the gap between simulations and observations persists. 

\BH{To quantify this, we can once again define a ``fixed amplitude'' and ``free amplitude'' $\smash{{\rm SNR}_{\rm TNG300\mbox{-}1} \equiv (\chi^2_{\rm TNG300\mbox{-}1} - \chi^2_{\rm bf})^{\frac{1}{2}}}$, with the best-fitting curve as usual being Illustris-1. We find   $\smash{{\rm SNR}_{\rm TNG300\mbox{-}1}^{\rm fixed}} = 17$ and 19 for the Main and Extended samples, and  $\smash{{\rm SNR}_{\rm TNG300\mbox{-}1}^{\rm free}} \approx 3$ for both samples. Even in the very conservative ``free amplitude'' case, we find a preference for a more extended profile and hence larger feedback compared to TNG300-1.}

\subsection{Implications for lensing}
\BH{Given the fact that there appears to be a mismatch between the baryons and the dark matter distributions within a few virial radii, and that baryons amount for $\sim$ 16\% of the mass, our findings imply that baryons have a non-negligible effect on lensing, especially on small scales. While we leave a detailed study to future work, we note that with a similar baryon distribution around the BOSS CMASS galaxies as measured by kSZ \citep{Schaan21}, the authors were able to show a $\sim$ 15\% suppression on the lensing signal on scales smaller than 1 Mpc, and a few percent suppression up to several Mpc \citep{2021PhRvD.103f3514A}. This is consistent with evacuating the majority of the baryons from sub-Mpc scales, leading to a suppression in the lensing signal of order the baryon mass fraction, and significantly mitigating the ``Lensing is low'' \cite{2017MNRAS.467.3024L} problem. We expect similar behavior here.}

\BH{On larger scales, one may wonder if baryons play a part in the ``$S_8$ tension'' \cite{2022JHEAp..34...49A} as measured by several weak lensing surveys: the impact will likely depend on the analysis choices of each specific experiment and their scale cuts, which are typically determined using lower-feedback simulations. For example, a joint analyses of DES data with the BOSS kSZ measurements hint that baryon effects can mitigate part of the tension \cite{2024arXiv240406098B, McCarthy:2024tvp}. Given that the measurements presented here are twice as precise as those used in previous work and cover a broader redshift range, similar analyses should be able to significantly strengthen the conclusions. We leave a detailed study to upcoming work.}

\section{Summary and implications}
\label{sec:conc}

In this paper, we present the highest SNR measurement of the gas profiles around galaxy groups using the kSZ effect: we detect the signal at 13$\sigma$ \BH{and find strong evidence of deviation from dark matter profiles, as well as a preference for larger feedback compared to some commonly used state-of-the-art hydrodynamical simulations such as IllustrisTNG.}

It also suggests that baryons might play a more significant role than assumed in many cosmological analyses and alleviate tensions such as `Lensing is low’ and `Low $S_8$.’ Properly accounting for baryonic feedback is critical to placing robust constraints on many open questions, including the mass of neutrinos and the nature of dark matter – questions that will be crucial to future cosmological surveys such as \textit{Roman}, \textit{Euclid}, and the Vera Rubin Observatory. Our measurements of the gas profiles can be used to calibrate the subgrid models of hydrodynamical simulations, a task typically complicated by the low detection sensitivity to gas on the outskirts of halos (i.e., the `missing baryon’ problem). Combining kSZ measurements with tSZ and X-ray measurements, which provide access to additional quantities such as the temperature and cooling rate around galaxies \citep{2017JCAP...11..040B}, will enable us to fully solve the gas thermodynamics of groups and clusters and shed light on the role of feedback in galaxy evolution.

To make the kSZ measurement, we use the ACT temperature map and the DESI photometric galaxy sample of LRGs, which we split into four redshift bins. We detect the signal with respect to the dark matter in each of them at $\gtrsim$10$\sigma$ (see Table~\ref{tab:chi2}). This allows us to study for the first time the redshift evolution of the baryonic feedback through kSZ stacking and place tighter constraints on the allowed astrophysical feedback models. The fact that we see little evolution of the signal suggests that the population of red galaxies probed by DESI is fairly stable and that the AGN feedback is of similar magnitude across $z = 0.4 \sim 1$. A major advantage of the gas profiles obtained using the kSZ is that they are practically systematics-free, as the velocity-weighted stacking we perform cancels additive contaminants such as the tSZ and CIB (see App.~\ref{app:null} on null tests as well as \citep{2016PhRvD..93h2002S}). Importantly, unlike many other astrophysical probes, the signal is directly proportional to the amount of gas and independent of other properties such as temperature.
To put our findings into perspective, we compare the observed gas density profiles with mock measurements extracted from the state-of-the-art hydrodynamical simulation TNG300-1 and its predecessor, Illustris-1. In particular, we mimic the stacking and LRG selection process of the observational analysis and test various scenarios related to the simulation targeting choices (satellite fraction, number density, halo mass, velocity uncertainty) to quantify the allowed variations (see Fig.~\ref{fig:kSZ_sim}). We find that the baryonic feedback in the TNG300-1 simulation is not sufficiently strong to push enough of the gas out of the halo virial radius, whereas Illustris-1 accomplishes that more successfully. Future work examining the origin of this is instrumental to reconciling the theory and observations of the gas-dark matter link.

As the first study of the kSZ signal measured around a photometric sample of galaxies with reconstructed velocities, this work opens the door for an exciting new line of research with imaging surveys such as the large-scale projects Vera Rubin Observatory and \textit{Euclid}, which will provide not only a much larger number of objects compared with their spectroscopic counterparts, but also more diverse samples that cover a wider range of redshifts, masses, morphologies, colors and environments. Understanding the connection between gas and dark matter will not only aid future cosmology analyses, but also help our understanding of galaxy formation and evolution. This paper adds an essential piece to a growing body of works aiming to unravel the complexities of cosmic gas in the era of large cosmological surveys.

\acknowledgements
Data points for the figures are available at \url{https://doi.org/10.5281/zenodo.12633573}.

We would like to thank Minh Nguyen, Vid Irsic, Alex Amon, Daniel Gruen, Yulin Gong, as well as the anonymous referees for providing very useful comments.

B.H. thanks the Miller Institute for financially supporting her postdoctoral research. S.F. is supported by Lawrence Berkeley National Laboratory and the Director, Office of Science, Office of High Energy Physics of the U.S. Department of Energy under Contract No.\ DE-AC02-05CH11231.
B.R. and E.S. received support from the U.S. Department of Energy under contract number DE-AC02-76SF00515 to SLAC National Accelerator Laboratory. This research used resources of the National Energy Research Scientific Computing Center (NERSC), a U.S. Department of Energy Office of Science User Facility located at Lawrence Berkeley National Laboratory, operated under Contract No. DE-AC02-05CH11231. The Flatiron Institute is supported by the Simons Foundation. EC acknowledges support from the European Research Council (ERC) under the European Union’s Horizon 2020 research and innovation programme (Grant agreement No. 849169). MM acknowledges support from NSF grants AST-2307727 and  AST-2153201 and NASA grant 21-ATP21-0145. GSF acknowledges support through the Isaac Newton Studentship and the Helen Stone Scholarship at the University of Cambridge. GSF and FQ furthermore acknowledge support from the European Research Council (ERC) under the European Union’s Horizon 2020 research and innovation programme (Grant agreement No. 851274). CS acknowledges support from the Agencia Nacional de Investigaci\'on y Desarrollo (ANID) through Basal project FB210003. CS acknowledges support from the Agencia Nacional de Investigaci\'on y Desarrollo (ANID) through Basal project FB210003. RH acknowledges funding from the NSERC Discovery Grant RGPIN-2018-05750 and the Connaught Fund. This research was enabled in part by support provided by the SciNet cluster and the Digital Research Alliance of Canada. This work was supported by a grant from the Simons Foundation (CCA 918271, PBL).

This material is based upon work supported by the U.S. Department of Energy (DOE), Office of Science, Office of High-Energy Physics, under Contract No. DE–AC02–05CH11231, and by the National Energy Research Scientific Computing Center, a DOE Office of Science User Facility under the same contract. Additional support for DESI was provided by the U.S. National Science Foundation (NSF), Division of Astronomical Sciences under Contract No. AST-0950945 to the NSF’s National Optical-Infrared Astronomy Research Laboratory; the Science and Technology Facilities Council of the United Kingdom; the Gordon and Betty Moore Foundation; the Heising-Simons Foundation; the French Alternative Energies and Atomic Energy Commission (CEA); the National Council of Humanities, Science and Technology of Mexico (CONAHCYT); the Ministry of Science, Innovation and Universities of Spain (MICIU/AEI/10.13039/501100011033), and by the DESI Member Institutions: \url{https://www.desi.lbl.gov/collaborating-institutions}. Any opinions, findings, and conclusions or recommendations expressed in this material are those of the author(s) and do not necessarily reflect the views of the U. S. National Science Foundation, the U. S. Department of Energy, or any of the listed funding agencies.

The authors are honored to be permitted to conduct scientific research on Iolkam Du’ag (Kitt Peak), a mountain with particular significance to the Tohono O’odham Nation.

The DESI Legacy Imaging Surveys consist of three individual and complementary projects: the Dark Energy Camera Legacy Survey (DECaLS), the Beijing-Arizona Sky Survey (BASS), and the Mayall $z$-band Legacy Survey (MzLS). DECaLS, BASS and MzLS together include data obtained, respectively, at the Blanco telescope, Cerro Tololo Inter-American Observatory, NSF's NOIRLab; the Bok telescope, Steward Observatory, University of Arizona; and the Mayall telescope, Kitt Peak National Observatory, NOIRLab. NOIRLab is operated by the Association of Universities for Research in Astronomy (AURA) under a cooperative agreement with the National Science Foundation. Pipeline processing and analyses of the data were supported by NOIRLab and the Lawrence Berkeley National Laboratory. Legacy Surveys also uses data products from the Near-Earth Object Wide-field Infrared Survey Explorer (NEOWISE), a project of the Jet Propulsion Laboratory/California Institute of Technology, funded by the National Aeronautics and Space Administration. Legacy Surveys was supported by: the Director, Office of Science, Office of High Energy Physics of the U.S. Department of Energy; the National Energy Research Scientific Computing Center, a DOE Office of Science User Facility; the U.S. National Science Foundation, Division of Astronomical Sciences; the National Astronomical Observatories of China, the Chinese Academy of Sciences and the Chinese National Natural Science Foundation. LBNL is managed by the Regents of the University of California under contract to the U.S. Department of Energy. The complete acknowledgments can be found at \url{https://www.legacysurvey.org/}.



Support for ACT was through the U.S.~National Science Foundation through awards AST-0408698, AST-0965625, and AST-1440226 for the ACT project, as well as awards PHY-0355328, PHY-0855887 and PHY-1214379. Funding was also provided by Princeton University, the University of Pennsylvania, and a Canada Foundation for Innovation (CFI) award to UBC. ACT operated in the Parque Astron\'omico Atacama in northern Chile under the auspices of the Agencia Nacional de Investigaci\'on y Desarrollo (ANID). The development of multichroic detectors and lenses was supported by NASA grants NNX13AE56G and NNX14AB58G. Detector research at NIST was supported by the NIST Innovations in Measurement Science program. Computing for ACT was performed using the Princeton Research Computing resources at Princeton University, the National Energy Research Scientific Computing Center (NERSC), and the Niagara supercomputer at the SciNet HPC Consortium. SciNet is funded by the CFI under the auspices of Compute Canada, the Government of Ontario, the Ontario Research Fund–Research Excellence, and the University of Toronto. We thank the Republic of Chile for hosting ACT in the northern Atacama, and the local indigenous Licanantay communities whom we follow in observing and learning from the night sky.

\appendix

\begin{figure}
    \centering
    \includegraphics[width=0.5\textwidth]{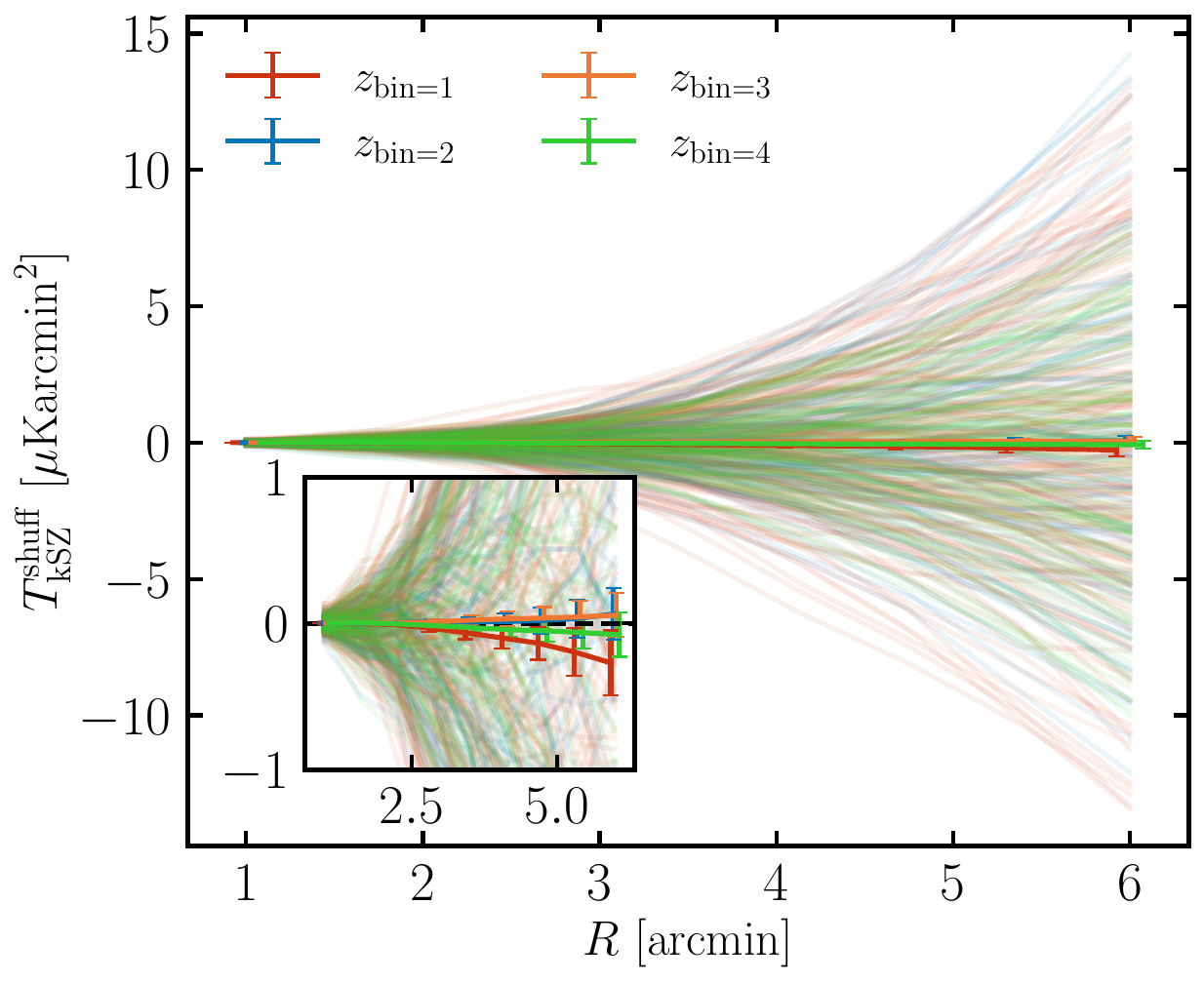}
    \caption{Null test demonstrating that there are no substantial systematics affecting our measurement such as residual CIB or tSZ contamination. The result is obtained by randomly shuffling the reconstructed velocities at the location of each DESI LRG before performing the stacking of the signal. The set of 1000 random reshuffles is shown as faint lines with its mean and error on the mean as solid lines. Reassuringly, it is consistent with zero and the largest deviations, $\sim$1, are much smaller ($\sim$10 times) than the size of the signal (Fig.~\ref{fig:kSZ}).
    }
    \label{fig:kSZ_null}
\end{figure}

\section{Null test}
\label{app:null}

In Fig.~\ref{fig:kSZ_null}, we validate that our measured signal is indeed due to the kSZ effect rather than correlated contaminants such as the cosmic infrared background (CIB) or tSZ. For this test, we shuffle randomly the sample of reconstructed velocities to obtain 1000 realizations
and perform the stacking of Eq.~\ref{eq:kSZ_est} but using the shuffled velocities on each galaxy. Since the true kSZ effect is proportional to the line-of-sight velocity, deleting that information in the shuffling should lead to a null measurement, which is indeed what we find. In particular, the null $\chi^2$ for the four redshift bins is 5.9, 5.0, 3.3 and 5.7, and the $p$-value is 0.75, 0.83, 0.95, 0.77, respectively. The tSZ and CIB effects do not have such dependence\footnote{Foreground emission such as from the CIB is also Doppler-boosted and acquires a term proportional to velocity. However, we find that this contribution is about two orders of magnitude smaller than the kSZ signal \cite{Maniyar:2022lkv}, and we'll therefore neglect it in this analysis.}, and hence if our ACT map is correctly cleaned, any additive contamination should also vanish. We note that this argument also holds true for the unshuffled (original) kSZ measurement, which is robust to additive contributions.

\section{Covariance matrix}
\label{app:cov}

\begin{figure}[ht]
    \centering
    \includegraphics[width=0.4\textwidth]{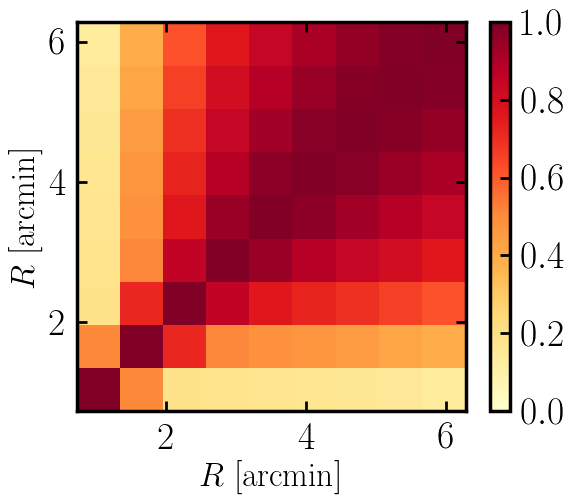}
    \caption{Correlation matrix of the kSZ signal between the different CAP filter radii for the first photometric bin. 
    The correlation between different CAP radii is stronger on large scales, as the fluctuations of the primary CMB become dominant on these scales. We note that the structure of the correlation function is the same for all samples.
    }
    \label{fig:kSZ_corr}
\end{figure}

\BH{The covariance matrix of the data is estimated via bootstrapping. In particular, for a given sample (e.g., the first redshift bin of the Main LRG sample), we generate 10,000 sets of kSZ measurements, where each set consists of a random draw (with repetition) from the kSZ measurements (equal to the number of galaxies) in that sample. We then compute the mean profile of each set and compute the covariance matrix across all 10,000 sets. We opt for 10,000 sets, as the error on the measurement error scales as the square root of that, and is thus about 1\%.}
This produces an unbiased estimate of the covariance, in the limit of independent noise realizations from galaxy to galaxy. The assumption is not correct for large apertures where the temperature map cutouts (submaps) overlap, and the inferred covariance is only accurate at the 10\% level (see \cite{Schaan21}). However, given our uncertainty, this level of accuracy is sufficient. 

The covariance matrix for the first photometric bin is shown in Fig.~\ref{fig:kSZ_corr}.
On small scales, the covariance is dominated by the detector noise in the temperature maps. Because this noise is mostly white and uncorrelated across frequencies, the small-aperture measurements are mostly uncorrelated within each submap and across submaps. On large scales, the covariance receives a large contribution from the primordial CMB fluctuations, which are shared between the aperture measurements in each submap and across submaps (and different CMB frequency maps). Due to this effect, which leads to diminishing returns in the SNR at larger apertures, the maximum aperture we consider is 6 arcmin.

\begin{figure}[ht]
    \centering
    \includegraphics[width=0.45\textwidth]{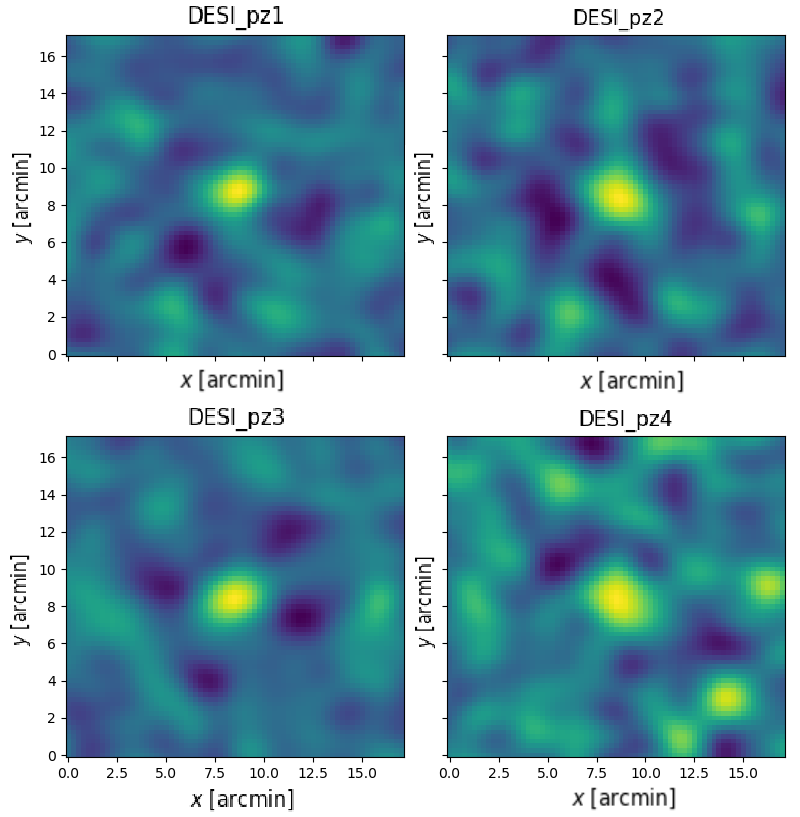}
    \caption{2D maps of the stacked kSZ signal around DESI DR9 LRGs in the Main sample for all four photometric bins. For visual purposes only, we high-pass filter the CMB temperature map before performing the stacking, as the fluctuations of the primary CMB are dominant in the $\sim$10 arcmin regime. Thus, we can clearly see the gas envelope of the DESI LRGs, which as expected, extends for several arcmin, i.e., is of the same order of magnitude as the mean halo virial radius at that redshift. \BH{Taking into account the mean virial radius of the galaxy groups (1.75 arcmin from TNG300) and the size of the CMB beam (1.6 arcmin), the beam-convolved mean virial radius can be estimated to be $\sqrt{((1.6^2/(8 \ln 2)+1.75^2} \approx 1.88$ arcmin.}}
    \label{fig:kSZ_stacks}
\end{figure}

In Fig.~\ref{fig:kSZ_stacks}, we illustrate the stacked 2D kSZ signal for the Main DR9 sample of DESI LRGs. We apply a Wiener filter ($C_\ell^{\rm kSZ}/C_\ell^{\rm tot}$) that effectively high-pass filters the CMB temperature map to isolate the small-scale signal that is the most affected by the kSZ effect. We can see by eye the extended gas envelope of the DESI LRGs, which spans several arcmin. \BH{In particular, using TNG300 we extract the average virial radius of the DESI LRGs at the mean redshift of the sample, 1.75 arcmin. As the CMB map is convolved with a 1.6 arcmin (FWHM) Gaussian beam, we can estimate the beam-convolved mean virial radius to be $\sqrt{((1.6^2/(8 \ln 2)+1.75^2} \approx 1.88$ arcmin.} This roughly corresponds to the mean halo virial radius at that redshift.

\section{Simulations}
\label{app:sims}

{To put our findings into context, we compare the gas density anisotropy we observe in the data against two hydrodynamical simulations: IllustrisTNG \citep{2019ComAC...6....2N} and Illustris \citep{2015A&C....13...12N}, which have very different AGN and supernova feedback prescriptions. To make the measurements in the simulations, we construct 2D maps of kSZ, then convolve them with a 1.6 arcmin Gaussian (corresponding to the DR6 map beam) to roughly match the ACT beam effects, 
and finally perform the stacking in an equivalent fashion as done to the data.}



{Each simulation outputs a number of useful quantities: gas mass ($m$), electron abundance ($x$), and internal energy ($\epsilon$) for each gas particle $i$. Assuming a primordial hydrogen mass fraction of $X_H=0.76$, we compute the volume-weighted electron number density, $n_{\rm e}$ as
\begin{equation}
V_i n_{{\rm e},i} = x_i m_i \frac{X_H}{m_p}
\end{equation}
where $m_p$ the proton mass \BH{and $V_i$ the volume of the gas cell}. We then compute the 2D maps of the kSZ and the optical depth by binning the gas particles into a (10000, 10000) grid, so that the optical depth in cell $j$ is given by:
\begin{equation}
\tau_j = \sigma_T A_j^{-1} \sum_{i \in A_j} V_i n_{{\rm e},i} ,
\end{equation}
and the momentum of the electron density is:
\begin{equation}
b_j = \sigma_T A_j^{-1} \sum_{i \in A_j} V_i n_{{\rm e},i} v_i/c,
\end{equation}
where $A_j$ is the area of each grid cell (of size $\sim$0.01 Mpc), $\sigma_T$ is the Thomson cross section, and $c$ is the speed of light. Analogously, we can compute the Y-Compton map relevant to the tSZ effect and the Thomson optical depth $\tau$ map relevant to e.g., patchy screening.
}

{We select LRG-like galaxies in the two simulations via an abundance matching approach and stack them at their locations on 2D CMB data mimicking ACT maps.
Namely, we rank-order the galaxies by stellar mass and select the top $N_{\rm gal}$ such that, $N_{\rm gal}/L^3 \sim 5.4 \times 10^{-4} \ [{\rm Mpc}/h]^{-3}$ for the Main sample and $1.2 \times 10^{-3} \ [{\rm Mpc}/h]^{-3}$ for the Extended sample. We find that the mean halo mass of LRGs in TNG300-1 matches very well the inferred mean halo mass of DESI LRGs at $0.4 < z < 0.6$, $10^{13.4} \ {\rm M}_\odot/h$, and at $0.6 < z < 0.8$, $10^{13.2} \ {\rm M}_\odot/h$ \citep{Yuan:2023ezi}. We then perform 2D stacking on the optical depth maps at the positions of these galaxy samples.}

\section{\BH{Effect of photometric noise and recovery of the true $\tau$}}

\BH{In this appendix, we explore two important effects and show that they are properly accounted for in our analysis. First, we study the effect of reconstruction on the shape of the CAP kSZ profiles. To this end, we repeat the CAP kSZ measurement detailed in Section~\ref{sec:methods} on an LRG-like galaxy sample, where we a) use the true host halo velocities in the stacking and b) use the reconstructed velocities with 2\% uncertainties on the photometric redshifts (similar to DR9 and DR10 used in this analysis). As in the case of the DESI data, we perform reconstruction by adopting the package \texttt{pyrecon}. Our simulation mocks aim to mimic the DESI LRG selection and the ACT CMB map as closely as possible. In particular, we employ the full-sky \textsc{AbacusSummit} simulation, \texttt{AbacusSummit\_huge\_c000\_ph201}, and populate its halo light cones with an HOD matching the properties of the Extended LRG sample (for more details, see Refs.~\citep{2022MNRAS.509.2194H,2024PhRvD.109j3534H}). We then add redshift space distortions to each galaxy and photometric noise of 2\%. We cut the galaxy catalog into the DESI footprint and perform reconstruction using the same settings in \texttt{pyrecon} as in the real data.}

\BH{On the CMB side, we generate the optical depth field on the \textsc{AbacusSummit} density shells utilizing the `Transfer Function' method of Ref.~\citep{Sharma:2024kwj, 2025arXiv250411794L}. When creating the kSZ maps, we multiply each optical depth shell by the line-of-sight velocity shell before integrating (i.e., summing up the shells) along the line-of-sight. Finally, we reproject the HEALPiX map into the shape and footprint of the ACT DR6 map using the package \texttt{pixell}. We can now treat these two data sets (the galaxy catalog and the kSZ/$\tau$ map) as if they are `real data' and perform the corresponding kSZ analysis. We correct the reconstructed profile by $1/r$ normalizing the estimator in Eq. \ref{eq:kSZ_est} to account for the imperfect reconstruction, with $r = 0.3$, appropriate for the photometric redshift error adopted here \cite{2024PhRvD.109j3534H}. To put these findings into the context of the present work, we show the error bars from our main result, Fig.~\ref{fig:kSZ_sim}.}

\begin{figure}[ht]
    \centering
    \includegraphics[width=0.45\textwidth]{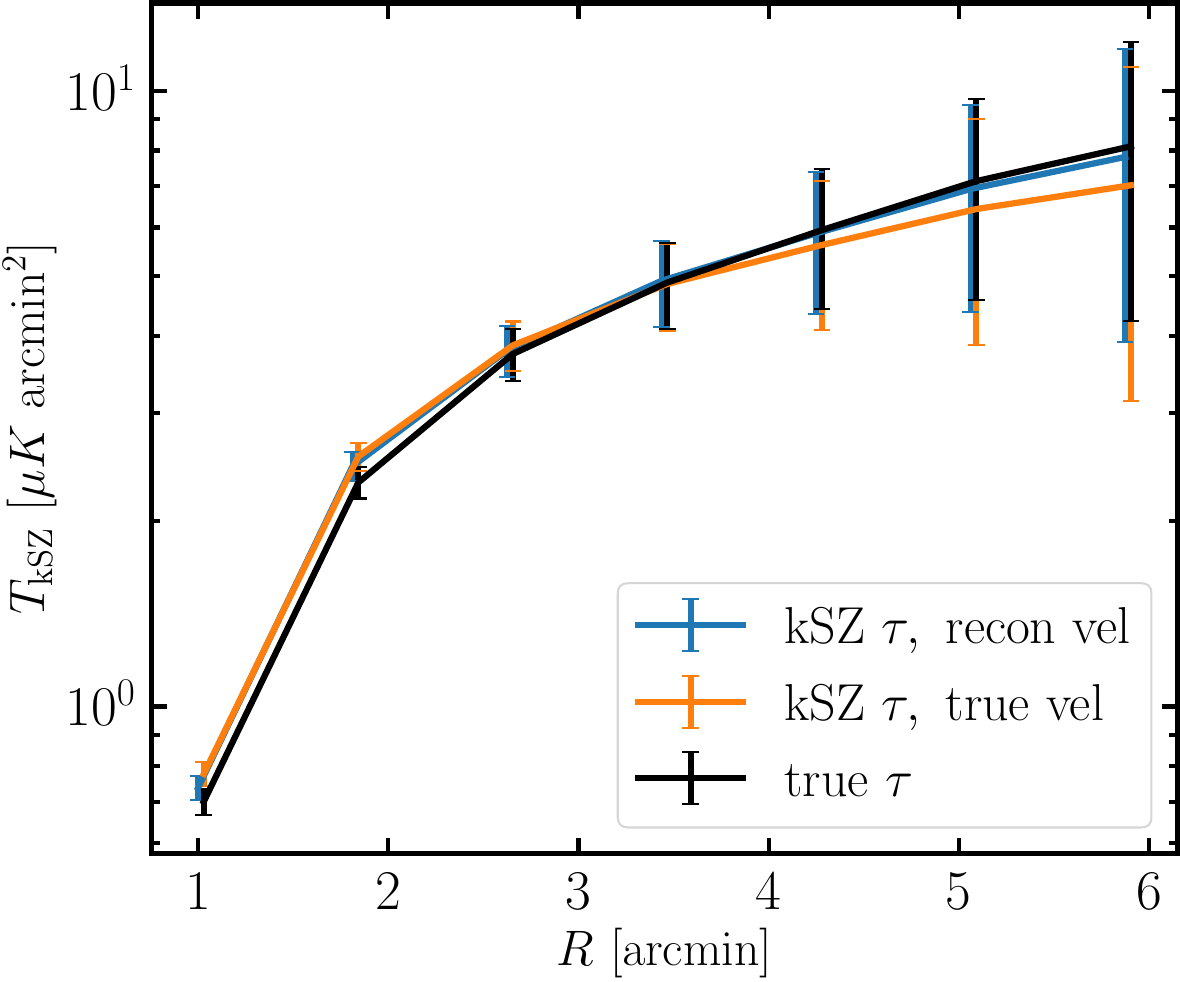}
    \caption{\BH{Comparison in simulations between the true input optical depth CAP profile and the inferred optical depth CAP profiles using either the true host halo velocity of each galaxy or the reconstructed velocities using the package \texttt{pyrecon}. In all three cases, the CAP profiles are measured around LRG-like galaxies, and when performing the reconstruction, we add a 2\% error on the photometric redshifts, akin to the photo-$z$ error of the LRG sample of DR9 and DR10. We correct the reconstructed profile by $1/r$ to account for the imperfect reconstruction, with $r = 0.3$. To evaluate the differences between these curves, we show the error bars from our data measurement (Fig.~\ref{fig:kSZ_sim}). All three curves are in very good agreement with each other and well within the measurement error bars. We find no evidence for scale dependence of the $r$ coefficient despite the presence of photometric noise and redshift space distortions, verifying our methodology. Moreover, the reconstructed curve recovers very well the true optical depth curve, lending further credence to our analysis. For details on the simulation, see the main text.}
    }
    \label{fig:tau_true_rec}
\end{figure}

\BH{As seen from Fig.~\ref{fig:tau_true_rec}, all three curves are in excellent agreement with each other and well within the measurement error bars. We find no evidence for scale dependence (0.1$\sigma$ difference between the two) of the $r$ coefficient despite the presence of photometric noise and redshift space distortions, verifying our methodology. This is expected because the cosmological velocity field has a correlation length much larger than a typical gas halo, and confirms the analytical results of \citep{Smith:2018bpn}. Furthermore, since the first version of this paper, the corresponding analysis for the DESI Y1 spectroscopic sample has appeared \citep{RiedGuachalla:2025byu}. A detailed comparison between the spectroscopic and photometric analyses is contained in that reference, but here we note that the excellent agreement between the two further confirms that photometric redshift errors do not bias the signal or introduce a scale dependence, beyond changing the value of $r$ as expected.}

\BH{The second takeaway from this plot is that the reconstructed curve recovers the true optical depth curve very well, demonstrating that our kSZ measurements indeed recover the true optical depth, as intended. }

\section{Mass evolution}
\label{app:mass}

\begin{figure}[ht]
    \centering
    \includegraphics[width=0.45\textwidth]{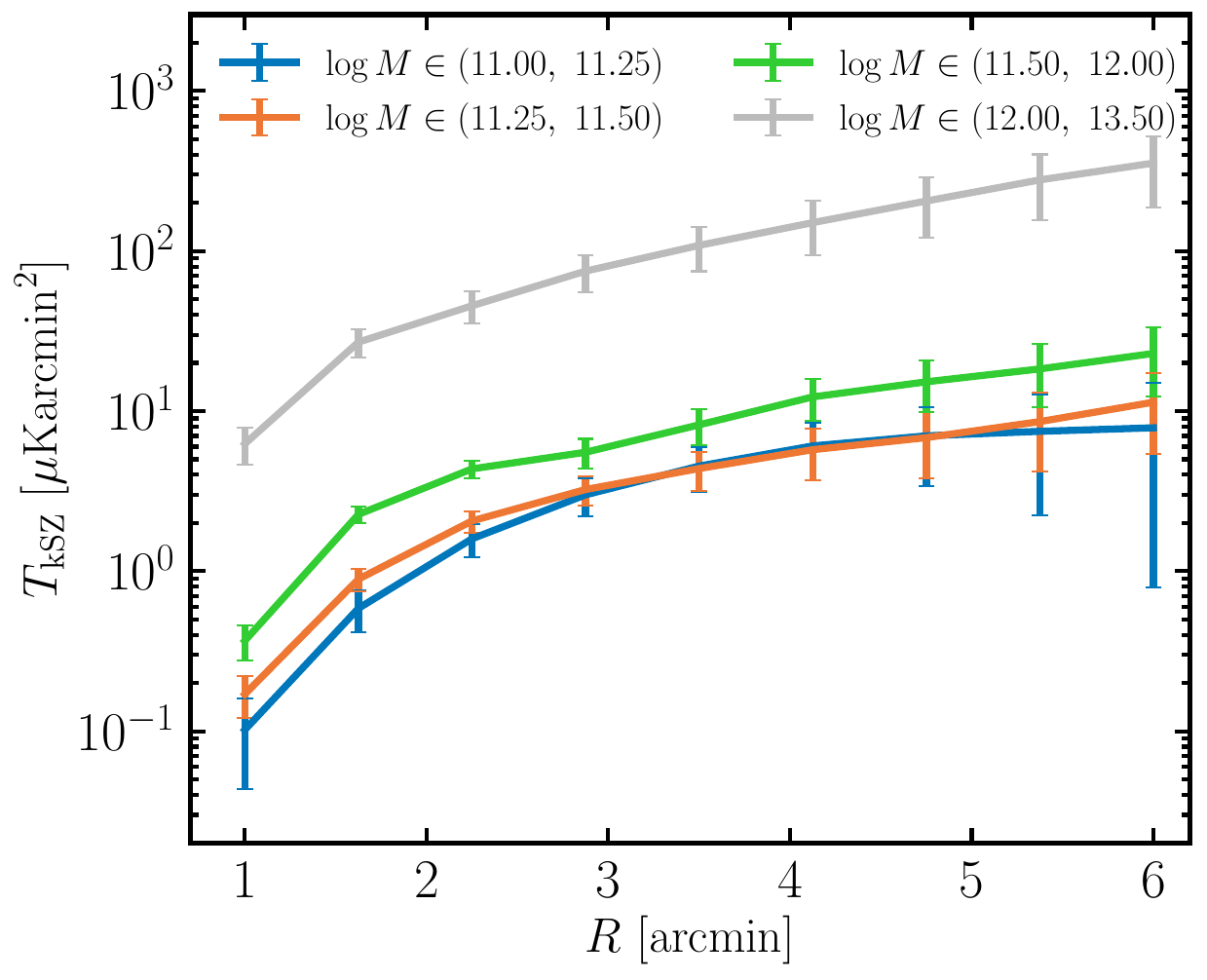}
    \caption{As Fig.~\ref{fig:kSZ_sim}, but we show the kSZ stacked measurements for the combined (DR9+DR10) Main sample, split into stellar mass bins. We see a noticeable increase in the amplitude of the signal with mass and note that the relative differences between these amplitudes are immune to miscalibration of e.g., the cross-correlation coefficient $r$, and the rms of the velocity. As expected, the curves become shallower at small apertures, as the average host halo mass increases.}
    \label{fig:kSZ_mass}
\end{figure}

Understanding the mass evolution of baryonic feedback is key to properly modeling its effect on cosmological observables such as cosmic shear, and the SZ effect is a versatile probe capable of capturing multiple gas properties \citep[see e.g.,][for novel probes and methods]{2021PhRvD.104d3518K,2021MNRAS.503.1798C,2023PhRvD.108b3516M,2023JCAP...03..039B,2024arXiv240218645M,2024arXiv240300909D}.
Here we use estimates of the stellar mass from \citet{2023AJ....165...58Z} to split the galaxies in the combined (DR9+DR10) Main sample into four mass/luminosity bins: $\log M \in (11, 11.25), (11.25, 11.5), (11.5, 12), \ {\rm and} \ (12, 13.5)$, each containing 926,232, 1,362,717, 458,886, and 2,152 galaxies, respectively. We note that the estimated masses have not been thoroughly characterized, and thus, they are mostly used to make qualitative statements. 

We show the kSZ stacked measurements for each bin in Fig.~\ref{fig:kSZ_mass}. We detect the individual signals at high significance. The null $\chi^2$ is $\chi^2_{\rm null} = 29.7$, 68.2, 96.2, and 28.3.
As can be seen in the figure, we find a strong signature of the evolution of the amplitude with mass: namely, it increases with stellar mass/luminosity, as the optical depth is proportional to the total mass of the halo, i.e., the amplitudes of the stellar-mass bin samples differ roughly by the ratios of their respective mean stellar masses, since $\tau \propto M_{\rm halo} \propto M_*$ \BH{(this breaks down as we go to higher halo masses, as the stellar-to-halo mass ratio decreases). As the mean stellar masses in each bin are 1.3, 2.3, 5.0 and 35 $10^{11} M_\odot/h$, respectively, we see that this statement holds well. We note that the relative differences between the amplitudes are independent of the calibration of the cross-correlation coefficient $r$ and the rms of the velocity. We investigate this in detail in simulations of the full DESI data (matching footprint, redshift distribution and halo occupation distribution of the sample) and find that the assumption of scale and mass independence of $r$ holds very well: the largest discrepancy we find across all radial bins when testing both the scale and mass independence is $\lesssim$0.2$\sigma$. In addition, as expected, the curves become shallower at small apertures, as the mass increases. This is particularly well seen for the least massive and most massive stellar mass bins, which display the steepest and shallowest profiles, respectively.} This is the case because massive halos retain more gas within the virial radius, as they have deeper potential wells compared with their less massive counterparts. At large apertures, the points are extremely strongly correlated, and it is imprudent to draw any definite conclusions on this issue (see App.~\ref{app:cov}). We leave a detailed analysis of the effect of baryons on weak lensing probes, as measured in this paper, to future work.

\section{Comparison with CMASS measurement}
\label{app:cmass}

\begin{figure}[h]
    \centering
    \includegraphics[width=0.47\textwidth]{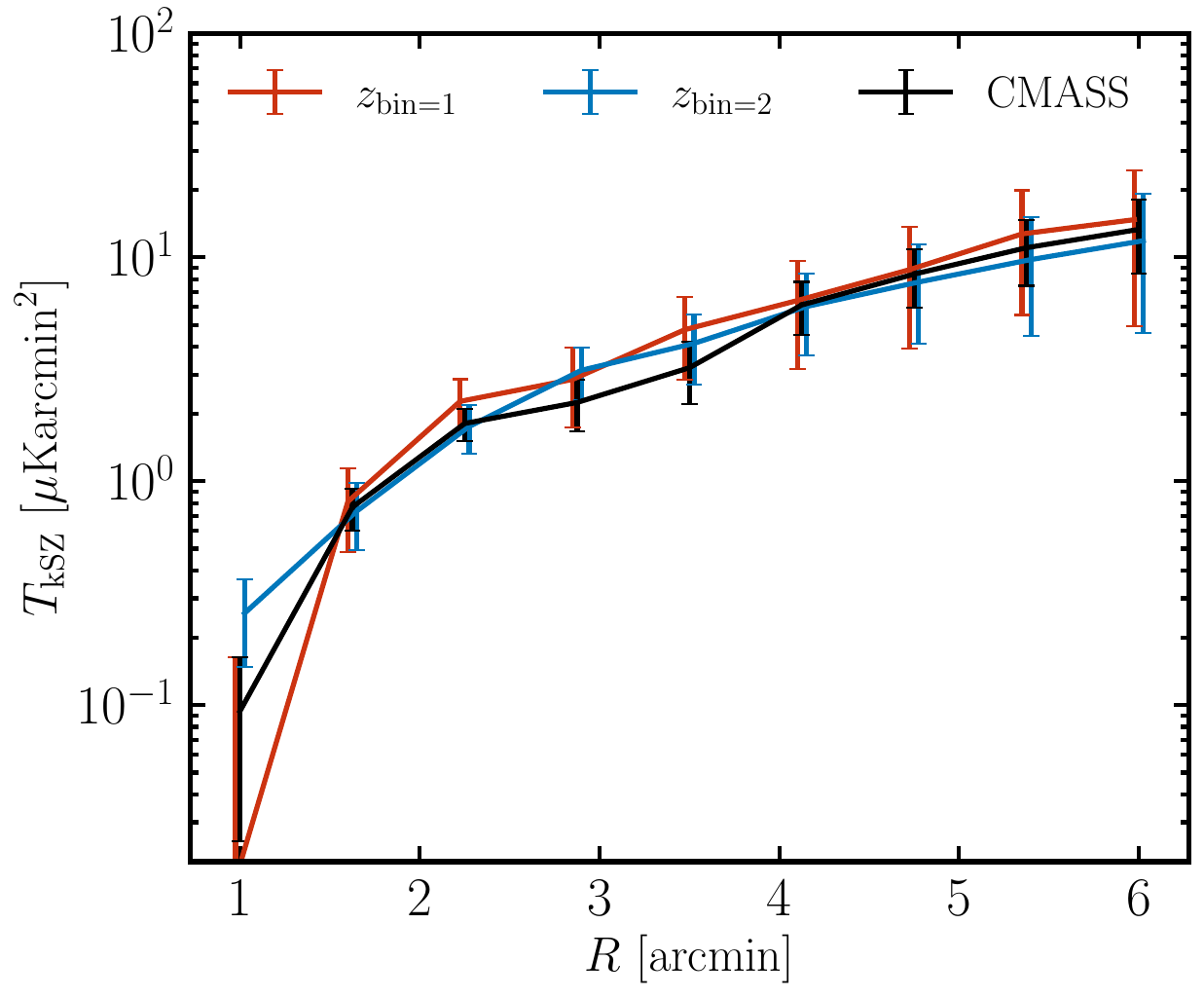}
    \caption{Comparison between the DESI photo-$z$ stacked kSZ signal in the first two redshift bins and the CMASS$\times$ACT stacked kSZ measurement from \cite{Schaan21}. Reassuringly, we see that the agreement with the first bin, which is closest in redshift to CMASS, $z_{\rm CMASS} \approx 0.55$ (cf. $z_{\rm bin=1} \approx 0.47$) is excellent. 
    To make this visual comparison at the same frequency, we use the DR5 ACT maps at 90 GHz.
    }
    \label{fig:kSZ_CMASS}
\end{figure}

In Fig.~\ref{fig:kSZ_CMASS}, we show a comparison between the DESI photo-$z$ stacked kSZ signal in each redshift bin and the old CMASS$\times$ACT stacked kSZ measurement from \cite{Schaan21}. We find excellent agreement between the two, in particular for the first redshift bin of our DESI result, which is closest in redshift to CMASS, $z_{\rm CMASS} \approx 0.55$ (cf. $z_{\rm bin=1} \approx 0.47$). We note that while the CMASS and the DESI LRG samples have similar host halo properties (e.g., mass, occupation distribution), they are not exactly the same, and therefore, this agreement should not be taken as a consistency check. 

We emphasize that the combined chi-squared 
of all four bins, $\chi^2_{\rm null} \approx 200$, is about twice higher than that for CMASS$\times$ACT, $\chi^2_{\rm null} \approx 86$. While the lower $r$ value for our photometric sample reduces the SNR by about half, the improvement in number of objects (by 10-20 times depending on the sample), more than compensates for the loss.

\section{Effect of removing corrections and comparison with Data Release 9}
\label{app:corr_dr9}

\begin{figure}[h]
    \centering
    \includegraphics[width=0.47\textwidth]{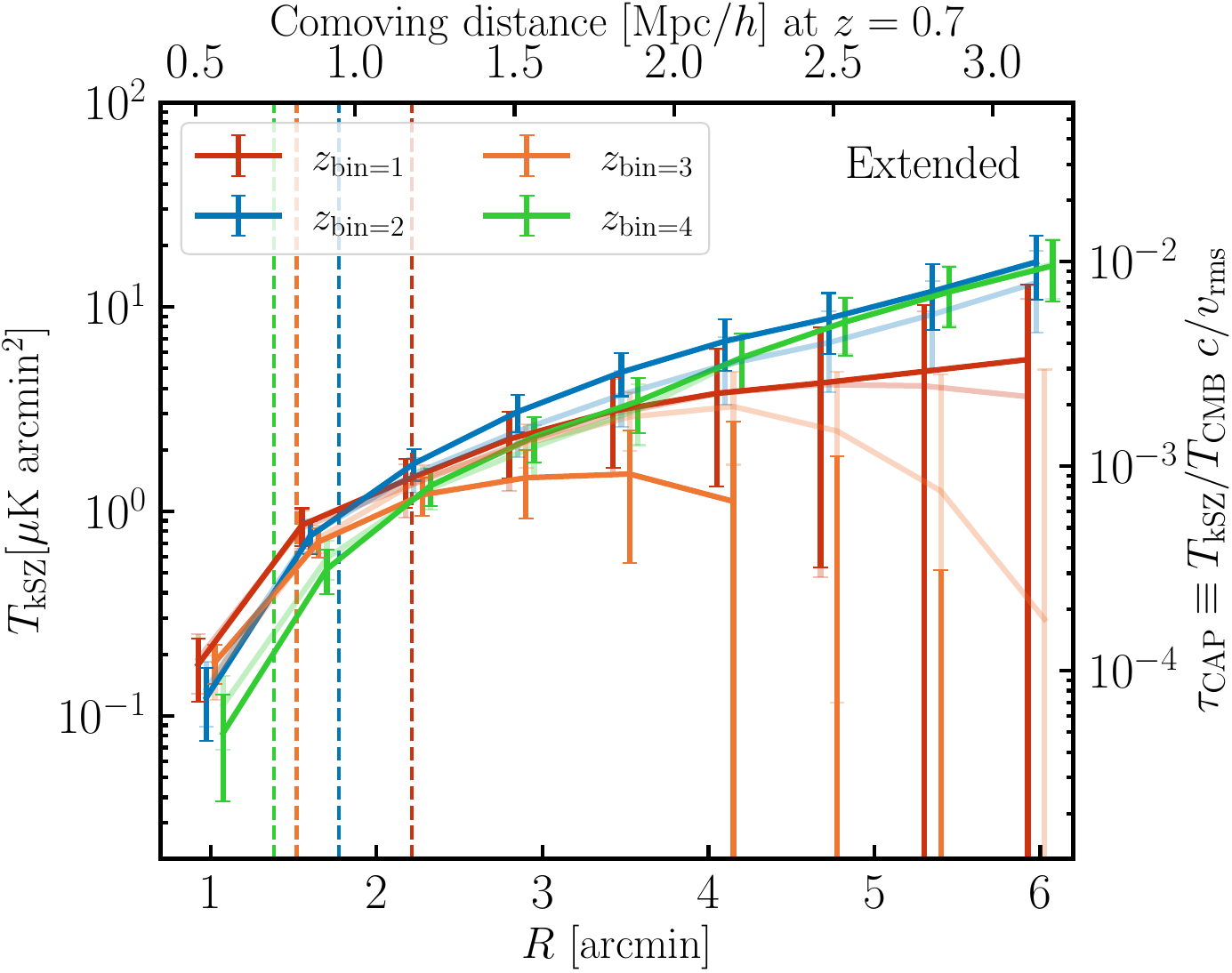}
    \includegraphics[width=0.47\textwidth]{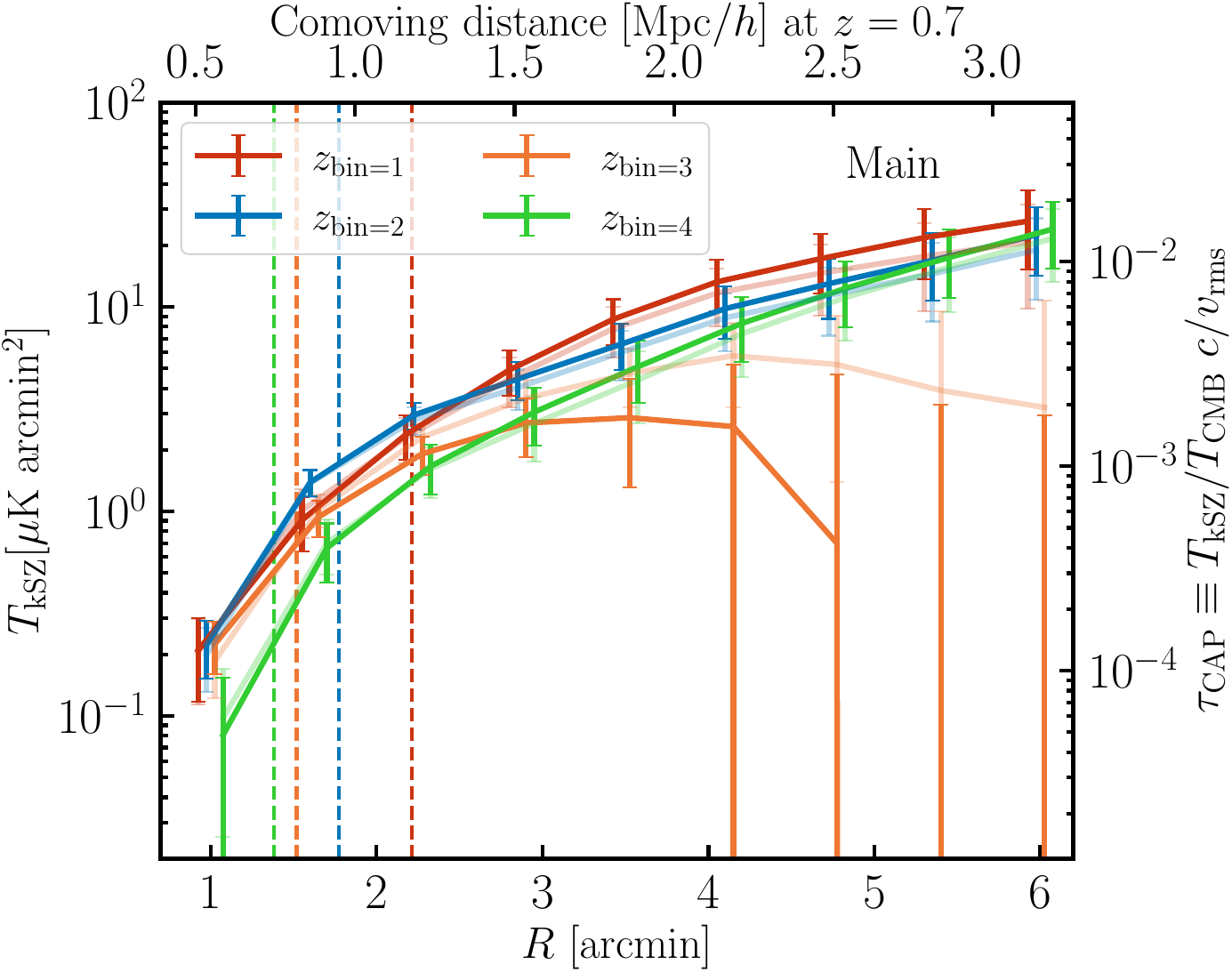}
    \caption{Same as ~\ref{fig:kSZ}, but shown for the case where no corrections are applied. The fiducial measurement is shown as faded curves. We see that the two cases are in very good agreement.}
    \label{fig:kSZ_nocorr_nosigz}
\end{figure}

As mentioned in the main text of this manuscipt, our fiducial measurement involves two additional cleaning techniques beyond the analysis of \cite{Schaan21}: namely, removing photometric $z$ outliers with estimated noise above $\sigma_z/(1+z) > 0.05$ and removing reconstructed velocity outliers at 3$\sigma$. In Table~\ref{tab:chi2_corr}, we compare the significance of the detection in the scenario of not applying each and any of these corrections.

Reassuringly, the three statistics of interest to this study, $\chi^2_{\rm null}$, ${\rm SNR}_{\rm DM}$ and ${\rm SNR}_{\rm null}$ are very similar in all four cases: both corrections (fiducial), only the photometric one, only the outlier one, and neither one applied. In particular, for the most part we see that the other three cases tend to have very similar if not slightly higher $\chi^2_{\rm null}$ compared with our fiducial analysis. We attribute this to the fact that the fiducial case features the smallest number of galaxies, having had a number of outliers removed.

A notable exception is the third photometric bin, which behaves better in the fiducial case than all others. Upon further inspection, we see that it has a slightly asymmetric reconstructed velocity distribution (possibly due to poorly reconstructed regions as a result of masking and/or photometric $z$ miscalibration). Overall, the highest SNR is found in the case where only a photometric $z$ error cleaning has been applied. We also note that the tails of the redshift distribution appear to be the largest in this bin \citep{Zhou:2023gji}.

To corroborate the claim that the additional cleaning applied in this work has minimal effect on the profile shapes, in Fig.~\ref{fig:kSZ_nocorr_nosigz}, we present a comparison between the fiducial (cleaned) analysis and the case where neither of these corrections is applied. Visually, the curves bear a very strong resemblance with each other, all measurements being within 1$\sigma$ of each other. Most prominently, we see that the third bin behaves more physically in the corrected case (since we adopt a CAP filter, we expect the measured profiles to increase until all gas is encompassed within some aperture radius\footnote{\BH{In particular, in the unphysical case of an isolated halo, the profile would flatten out once all the gas is encompassed. However, due to the presence of other correlated structure nearby (i.e., two-halo term), we expect the profiles to keep increasing mildly with radius. In reality, on large scales, the noise is highly correlated and dominated by the large variance of the primary CMB.}}), while the signal in the first bin gets a slight boost when no corrections are applied. This is reassuring to see, as it suggests that our cleaning procedure works as expected (i.e., it does not significantly bias our outputs and reduces the anomalous behavior exhibited by the third bin).

Furthermore, in Table~\ref{tab:chi2_corr_dr9}, we examine the stacked kSZ measurement coming from the DR9 samples and juxtapose it with the fiducial DR10 result. Overall, the DR10 signal is higher than the DR9 one. This is particularly true for the higher-redshift bins, for which the DR9 photometric redshift estimates are significantly worse than DR10 and a consequence of the fact that the $i$-band colors are very powerful at constraining higher redshift. Interestingly, we find a stronger detection for the first two bins due to the larger number objects in DR9 (not all galaxies have $i$-band colors measured). Similarly to the DR10 case, we find that often the uncorrected analysis yields a higher-significance detection but for the third bin. Once again the photo-$z$ corrected sample boasts with some of the highest $\chi^2_{\rm null}$ values across all redshifts.

\newpage
    \begin{table}
        \small
        \centering
        {\renewcommand{\arraystretch}{1.5}
        \begin{tabular}{l||c|c|c|c|c}
            \hline
            \hline
            LRG sample & \# of galaxies & $z_{\rm mean}$ & $\chi^2_{\rm null}$ & ${\rm SNR}_{\rm DM}$ & ${\rm SNR}_{\rm null}$ \\[-0.5pt] 
            \hline
        \multicolumn{6}{c}{\textbf{DR10: Outlier and photo-$z$ correction (fiducial)}} \\[-0.5pt]
        Extended $z_{\rm bin=1}$ & 856,537 & 0.47 & 27.5 & 15.5 & 5.0 \\[-0.5pt]
        Extended $z_{\rm bin=2}$ & 1,422,411 & 0.63 & 49.0 & 19.1 & 6.0 \\[-0.5pt]
        Extended $z_{\rm bin=3}$ & 1,951,646 & 0.79 & 80.3 & 26.1 & 8.2 \\[-0.5pt]
        Extended $z_{\rm bin=4}$ & 1,690,171 & 0.92 & 34.0 & 14.6 & 4.6 \\[-0.5pt]
        \hline
        Extended all & 5,931,939 & 0.75 & 151.2 & 36.9 & 11.6 \\[-0.5pt]
        \hline
        Main $z_{\rm bin=1}$ & 374,555 & 0.47 & 23.7 & 14.1 & 4.3 \\[-0.5pt]
        Main $z_{\rm bin=2}$ & 668,450 & 0.63 & 62.2 & 22.6 & 7.1 \\[-0.5pt]
        Main $z_{\rm bin=3}$ & 753,945 & 0.79 & 65.9 & 23.4 & 7.3 \\[-0.5pt]
        Main $z_{\rm bin=4}$ & 629,367 & 0.93 & 20.8 & 10.1 & 3.2 \\[-0.5pt]
        \hline
        Main all & 2,438,749 & 0.73 & 126.2 & 33.8 & 10.5 \\[-0.5pt]
        \hline
        \multicolumn{6}{c}{\textbf{DR10: Photo-$z$ correction only}} \\[-0.5pt]
        Extended $z_{\rm bin=1}$ & 868,743 & 0.47 & 29.1 & 15.8 & 5.2 \\[-0.5pt]
        Extended $z_{\rm bin=2}$ & 1,435,540 & 0.63 & 57.3 & 20.9 & 6.5 \\[-0.5pt]
        Extended $z_{\rm bin=3}$ & 2,006,009 & 0.79 & 80.8 & 24.8 & 8.0 \\[-0.5pt]
        Extended $z_{\rm bin=4}$ & 1,704,265 & 0.92 & 35.1 & 15.2 & 4.8 \\[-0.5pt]
        \hline
        Extended all & 6,014,557 & 0.75 & 164.8 & 38.4 & 12.1 \\[-0.5pt]
        \hline
        Main $z_{\rm bin=1}$ & 380,052 & 0.47 & 27.0 & 14.8 & 4.5 \\[-0.5pt]
        Main $z_{\rm bin=2}$ & 677,185 & 0.63 & 74.0 & 24.4 & 7.7 \\[-0.5pt]
        Main $z_{\rm bin=3}$ & 771,391 & 0.79 & 60.7 & 21.7 & 6.9 \\[-0.5pt]
        Main $z_{\rm bin=4}$ & 636,964 & 0.93 & 20.4 & 10.4 & 3.2 \\[-0.5pt]
        \hline
        Main all & 2,465,593 & 0.73 & 145.4 & 36.0 & 11.2 \\[-0.5pt]
        \hline
        \multicolumn{6}{c}{\textbf{DR10: Outlier correction only}} \\[-0.5pt]
        Extended $z_{\rm bin=1}$ & 898,659 & 0.47 & 25.9 & 15.0 & 4.8 \\[-0.5pt]
        Extended $z_{\rm bin=2}$ & 1,476,128 & 0.63 & 48.9 & 18.8 & 5.9 \\[-0.5pt]
        Extended $z_{\rm bin=3}$ & 2,031,590 & 0.79 & 81.5 & 25.5 & 8.0 \\[-0.5pt]
        Extended $z_{\rm bin=4}$ & 1,753,100 & 0.92 & 26.6 & 13.2 & 4.0 \\[-0.5pt]
        \hline
        Extended all & 6,174,499 & 0.74 & 142.8 & 35.2 & 11.0 \\[-0.5pt]
        \hline
        Main $z_{\rm bin=1}$ & 393,180 & 0.47 & 19.7 & 12.8 & 3.8 \\[-0.5pt]
        Main $z_{\rm bin=2}$ & 690,663 & 0.63 & 58.2 & 20.5 & 6.5 \\[-0.5pt]
        Main $z_{\rm bin=3}$ & 775,891 & 0.79 & 70.4 & 22.9 & 7.1 \\[-0.5pt]
        Main $z_{\rm bin=4}$ & 645,219 & 0.93 & 20.9 & 9.8 & 3.0 \\[-0.5pt]
        \hline
        Main all & 2,515,452 & 0.73 & 120.7 & 31.9 & 9.9 \\[-0.5pt]
        \hline
        \multicolumn{6}{c}{\textbf{DR10: No corrections}} \\[-0.5pt]
        Extended $z_{\rm bin=1}$ & 911,407 & 0.47 & 30.2 & 16.6 & 5.3 \\[-0.5pt]
        Extended $z_{\rm bin=2}$ & 1,490,492 & 0.63 & 53.9 & 20.1 & 6.2 \\[-0.5pt]
        Extended $z_{\rm bin=3}$ & 2,083,198 & 0.79 & 80.8 & 24.2 & 7.8 \\[-0.5pt]
        Extended $z_{\rm bin=4}$ & 1,769,651 & 0.92 & 30.0 & 14.4 & 4.4 \\[-0.5pt]
        \hline
        Extended all & 6,254,748 & 0.74 & 156.4 & 37.0 & 11.7 \\[-0.5pt]
        \hline
        Main $z_{\rm bin=1}$ & 398,735 & 0.47 & 24.0 & 13.8 & 4.1 \\[-0.5pt]
        Main $z_{\rm bin=2}$ & 698,633 & 0.63 & 67.8 & 22.9 & 7.2 \\[-0.5pt]
        Main $z_{\rm bin=3}$ & 790,871 & 0.79 & 62.9 & 21.5 & 6.8 \\[-0.5pt]
        Main $z_{\rm bin=4}$ & 652,610 & 0.93 & 20.2 & 10.0 & 3.1 \\[-0.5pt]
        \hline
        Main all & 2,540,850 & 0.73 & 136.6 & 34.4 & 10.7 \\[-0.5pt]
        \hline
        \hline
            \end{tabular}%
            }
            \caption{Same as Table~\ref{tab:chi2}, but showing the effect of the different corrections applied to DR10 LRGs. The statistical significance of the measurements changes minimally.}
            \label{tab:chi2_corr}
        \end{table}

    \begin{table}
        \small
        \centering
        {\renewcommand{\arraystretch}{1.5}
        \begin{tabular}{l||c|c|c|c|c}
            \hline
            \hline
            LRG sample & \# of galaxies & $z_{\rm mean}$ & $\chi^2_{\rm null}$ & ${\rm SNR}_{\rm DM}$ & ${\rm SNR}_{\rm null}$ \\[-0.5pt] 
            \hline
        \multicolumn{6}{c}{\textbf{DR9: Outlier and photo-$z$ correction (fiducial)}} \\[-0.5pt]
        Extended $z_{\rm bin=1}$ & 954,820 & 0.47 & 38.5 & 18.7 & 6.0 \\[-0.5pt]
        Extended $z_{\rm bin=2}$ & 1,628,650 & 0.63 & 71.2 & 22.7 & 7.2 \\[-0.5pt]
        Extended $z_{\rm bin=3}$ & 2,125,787 & 0.80 & 35.6 & 14.2 & 4.2 \\[-0.5pt]
        Extended $z_{\rm bin=4}$ & 1,996,525 & 0.92 & 19.1 & 8.9 & 2.9 \\[-0.5pt]
        \hline
        Extended all & 6,697,792 & 0.75 & 98.7 & 29.3 & 9.1 \\[-0.5pt]
        \hline
        Main $z_{\rm bin=1}$ & 417,262 & 0.47 & 20.6 & 12.3 & 4.0 \\[-0.5pt]
        Main $z_{\rm bin=2}$ & 779,408 & 0.63 & 77.4 & 25.0 & 7.9 \\[-0.5pt]
        Main $z_{\rm bin=3}$ & 878,442 & 0.79 & 22.9 & 12.8 & 3.8 \\[-0.5pt]
        Main $z_{\rm bin=4}$ & 765,542 & 0.92 & 12.5 & 8.6 & 2.7 \\[-0.5pt]
        \hline
        Main all & 2,840,853 & 0.74 & 98.2 & 30.0 & 9.4 \\[-0.5pt]
        \hline
        \multicolumn{6}{c}{\textbf{DR9: Photo-$z$ correction only}} \\[-0.5pt]
        Extended $z_{\rm bin=1}$ & 963,631 & 0.47 & 43.2 & 19.9 & 6.4 \\[-0.5pt]
        Extended $z_{\rm bin=2}$ & 1,658,313 & 0.63 & 69.4 & 21.6 & 7.1 \\[-0.5pt]
        Extended $z_{\rm bin=3}$ & 2,174,053 & 0.80 & 28.5 & 10.9 & 3.4 \\[-0.5pt]
        Extended $z_{\rm bin=4}$ & 2,054,075 & 0.92 & 26.2 & 11.2 & 3.8 \\[-0.5pt]
        \hline
        Extended all & 6,850,072 & 0.75 & 111.7 & 30.1 & 9.7 \\[-0.5pt]
        \hline
        Main $z_{\rm bin=1}$ & 422,350 & 0.47 & 25.2 & 13.7 & 4.4 \\[-0.5pt]
        Main $z_{\rm bin=2}$ & 795,393 & 0.63 & 78.8 & 24.4 & 7.8 \\[-0.5pt]
        Main $z_{\rm bin=3}$ & 888,934 & 0.79 & 23.6 & 11.9 & 3.7 \\[-0.5pt]
        Main $z_{\rm bin=4}$ & 776,227 & 0.92 & 15.3 & 9.8 & 3.2 \\[-0.5pt]
        \hline
        Main all & 2,882,904 & 0.74 & 103.1 & 30.0 & 9.6 \\[-0.5pt]
        \hline
        \multicolumn{6}{c}{\textbf{DR9: Outlier correction only}} \\[-0.5pt]
        Extended $z_{\rm bin=1}$ & 1,065,110 & 0.47 & 37.1 & 18.4 & 5.9 \\[-0.5pt]
        Extended $z_{\rm bin=2}$ & 1,775,088 & 0.63 & 73.0 & 20.6 & 6.7 \\[-0.5pt]
        Extended $z_{\rm bin=3}$ & 2,331,881 & 0.80 & 32.8 & 14.3 & 4.2 \\[-0.5pt]
        Extended $z_{\rm bin=4}$ & 2,386,519 & 0.92 & 33.4 & 12.5 & 3.6 \\[-0.5pt]
        \hline
        Extended all & 7,596,154 & 0.75 & 92.3 & 28.0 & 8.6 \\[-0.5pt]
        \hline
        Main $z_{\rm bin=1}$ & 464,568 & 0.47 & 18.1 & 11.2 & 3.6 \\[-0.5pt]
        Main $z_{\rm bin=2}$ & 837,280 & 0.63 & 86.5 & 26.1 & 8.2 \\[-0.5pt]
        Main $z_{\rm bin=3}$ & 923,974 & 0.79 & 25.1 & 13.6 & 4.1 \\[-0.5pt]
        Main $z_{\rm bin=4}$ & 845,101 & 0.92 & 15.3 & 8.9 & 2.8 \\[-0.5pt]
        \hline
        Main all & 3,081,255 & 0.74 & 96.7 & 29.8 & 9.2 \\[-0.5pt]
        \hline
        \multicolumn{6}{c}{\textbf{DR9: No corrections}} \\[-0.5pt]
        Extended $z_{\rm bin=1}$ & 1,073,362 & 0.47 & 37.6 & 18.9 & 6.0 \\[-0.5pt]
        Extended $z_{\rm bin=2}$ & 1,804,919 & 0.63 & 73.2 & 21.0 & 7.0 \\[-0.5pt]
        Extended $z_{\rm bin=3}$ & 2,400,937 & 0.80 & 22.7 & 9.9 & 3.1 \\[-0.5pt]
        Extended $z_{\rm bin=4}$ & 2,485,696 & 0.92 & 32.3 & 14.8 & 4.9 \\[-0.5pt]
        \hline
        Extended all & 7,764,914 & 0.75 & 96.9 & 28.3 & 9.0 \\[-0.5pt]
        \hline
        Main $z_{\rm bin=1}$ & 470,133 & 0.47 & 20.6 & 12.3 & 4.0 \\[-0.5pt]
        Main $z_{\rm bin=2}$ & 854,005 & 0.63 & 85.2 & 25.0 & 8.0 \\[-0.5pt]
        Main $z_{\rm bin=3}$ & 936,270 & 0.79 & 22.2 & 11.4 & 3.6 \\[-0.5pt]
        Main $z_{\rm bin=4}$ & 857,753 & 0.92 & 14.8 & 9.3 & 2.9 \\[-0.5pt]
        \hline
        Main all & 3,118,161 & 0.74 & 95.6 & 28.9 & 9.1 \\[-0.5pt]
        \hline
        \hline
            \end{tabular}%
            }
            \caption{Same as Table~\ref{tab:chi2_corr}, but showing DR9 instead of DR10 LRGs. We see that the statistical significance is similar to the DR10 results, but slightly reduced.}
            \label{tab:chi2_corr_dr9}
        \end{table}

\bibliography{apssamp}

\end{document}